\documentclass{article}

\usepackage{PRIMEarxiv}
\usepackage{amssymb}
\usepackage[T1]{fontenc}
\usepackage{graphicx}
\usepackage{color}
\usepackage{color, soul}
\usepackage{microtype}
\soulregister\cite7
\soulregister\ref7
\soulregister\pageref7
\usepackage{multirow}
\usepackage{comment}
\usepackage{amssymb}
\usepackage{pifont}
\usepackage{amsmath} 
\usepackage{algorithm2e}
\usepackage{booktabs}
\usepackage{subcaption}
\usepackage{multirow}
\usepackage{makecell}  
\usepackage{dsfont}
\usepackage{url}
\usepackage{tikz}
\usetikzlibrary{arrows.meta, positioning, shapes, decorations.pathreplacing, calc, 3d}
\usepackage{pgfplots}
\pgfplotsset{compat=1.17}

\title{A Multi-task Mixture of Experts Framework for Malware Classification, Packing Detection, and Family Attribution}

\author{
Jithin S.\\
Department of Computer Applications, \\
Cochin University of Science \\
and Technology, India \\
\texttt{jithinmca@pg.cusat.ac.in} \\
\And
Roshin Sleeba C.\\
Department of Computer Applications, \\
Cochin University of Science \\
and Technology, India \\
\texttt{roshinsleebac2002@pg.cusat.ac.in} \\
\And
Anvin Mariya P. B.\\
Department of Computer Applications, \\
Cochin University of Science \\
and Technology, India \\
\texttt{anvinmariyapb@pg.cusat.ac.in} \\
\And
Asmitha K. A.\\
Department of Computer Applications, \\
Cochin University of Science \\
and Technology, India \\
\texttt{asmitha@pg.cusat.ac.in} \\
\And
Vinod P. \\
Department of Computer Applications, \\
Cochin University of Science \\
and Technology, India \\
\texttt{vinod.p@cusat.ac.in} \\
\And
Serena Nicolazzo \\
Department of Electrical, Computer and Biomedical Engineering,\\
University of Pavia, Italy\\
\texttt{serena.nicolazzo@unipv.it} \\
\And
Antonino Nocera\\
Department of Electrical, Computer and Biomedical Engineering,\\
University of Pavia, Italy\\
\texttt{antonino.nocera@unipv.it} \\
}

\begin{document}

\maketitle
\date{July 2026}

\begin{abstract}
Malware classification remains a challenging problem due to its inherent heterogeneity, the presence of packed binaries, and the diverse distribution of malware families. Traditional single-model detection mechanisms often fail to generalize across such diverse data, leading to degraded performance, particularly on obfuscated and rare malware samples. In this work, we propose a unified multi-task malware analysis framework based on Mixture of Experts (MoE) architectures. The proposed system evaluates performance across two different input representations, i.e., high-dimensional EMBER feature sets and raw 1D byte arrays extracted from Portable Executable files. It simultaneously performs three critical tasks: malware family classification, packed versus unpacked detection, and malware versus benign identification. By decomposing the problem into specialized expert networks and employing adaptive gating mechanisms, the model enables effective task-specific learning while maintaining overall scalability. We investigate multiple architectural variants, including Homogeneous MoE, Heterogeneous MoE, and Multi-Gate MoE (MMoE). Performance is evaluated in both standard and adversarial settings using original and mutated samples. The obtained results demonstrate that the Multi-Gate MoE model achieves the best performance, reaching a combined detection rate of 0.9744 with only $2.56\%$ failure rate. Moreover, this configuration exhibits improved robustness under mutation-induced distribution shifts. Our findings highlight the effectiveness of expert specialization and task-specific routing in handling complex malware distributions, making the proposed framework a promising direction for scalable and resilient malware detection systems.
\end{abstract}


\section{Introduction}

The rapid growth of digital infrastructure has been accompanied by an equally aggressive evolution of malware, making it a critical global concern. Modern malware is no longer limited to static and easily identifiable patterns. Instead, sophisticated techniques such as packing, encryption, and polymorphic transformations are often adopted to evade detection systems. As a result, traditional signature-based and heuristic-driven approaches are inadequate, particularly when dealing with previously unseen or obfuscated threats.

Malware analysis is typically approached through static and dynamic methods. Although static analysis offers efficiency by examining executable structures without execution, it struggles to handle heavily obfuscated or packed binaries. On the other hand, dynamic analysis provides richer behavioral insights but comes with significant computational overhead and practical deployment constraints. These limitations highlight the need for intelligent, scalable solutions that can adapt to diverse and evolving malware characteristics.

Recent advances in machine learning, particularly deep learning, have introduced new possibilities for malware detection. Feature-based representations such as EMBER enable the extraction of high-dimensional numerical descriptors from Portable Executable (PE) files, capturing structural, statistical, and semantic properties. However, a fundamental challenge remains: malware data is inherently heterogeneous. Variations between families, differences between packed and unpacked samples, and imbalances in class distributions make it difficult for a single monolithic model to generalize effectively across all scenarios.

To address these limitations, this work investigates the applicability of Mixture of Experts (MoE) architectures for malware analysis. MoE frameworks divide the learning process among multiple specialized expert networks, enabling different experts to focus on distinct malware characteristics and behavioral patterns. A gating mechanism dynamically selects and combines expert outputs based on the input sample, allowing the framework to adaptively model heterogeneous malware distributions. Such conditional computation is particularly beneficial in cybersecurity applications, where malware families, packing artifacts, and adversarial mutations often exhibit highly diverse representations. By allowing expert specialization, MoE architectures improve feature learning, enhance task-specific discrimination, and reduce interference between concurrent security objectives.

Therefore, in this paper, we propose a unified multi-task learning framework that simultaneously performs malware family classification, packed versus unpacked detection, and malware versus benign discrimination. We systematically investigate three variants of the MoE paradigm: Homogeneous MoE, Heterogeneous MoE, and Multi-Gate MoE (MMoE). Furthermore, we evaluate the robustness of these models under adversarial conditions by introducing mutation-based data augmentation.
Based on the challenges identified in multi-task malware analysis, this study seeks to address the following research questions:

\begin{itemize}
    \item \textbf{RQ1}: To what extent does the transition from a single-gate Mixture of Experts (MoE) to a Multi-Gate Mixture of Experts (MMoE) architecture improve performance in a multi-task malware analysis environment?
    
    \item \textbf{RQ2}: How do EMBER structured feature sets compare to raw 1D image representations in terms of classification accuracy and computational stability within an MoE framework?
    
    \item \textbf{RQ3}: How resilient are task-specific gating mechanisms against distribution shifts introduced by mutation-based metamorphic adversarial attacks compared to unified gating strategies?
\end{itemize}

With the objective of providing an answer to the previous research question, we hence derive the following contributions.

\begin{itemize}
\item We formulate malware analysis as a unified multi-task learning problem, integrating malware family classification, packing detection, and malware/ benign separation into a single framework.
\item We design and evaluate multiple Mixture of Experts architectures, highlighting the impact of expert diversity and gating strategies.
\item We perform a comprehensive comparative analysis between structured EMBER feature sets and raw 1D image representations.
\item We use mutation-based transformations to evaluate the model robustness under distribution shifts.
\item We demonstrate that Multi-Gate MoE achieves good performance and stability, particularly under distribution shifts introduced by mutated samples.
\end{itemize}

The results show that task-specific expert routing significantly improves both accuracy and generalization, making the proposed framework a strong candidate for future malware detection systems.

The remainder of this paper is organized as follows. Section \ref{sec:related} reviews the related work in malware analysis, deep learning, and Mixture of Experts architectures. Section \ref{sec:method} presents the proposed methodology, including the HOMO-MoE, HETERO-MoE, and Multi-Gate MoE frameworks along with the feature representations and learning strategies employed. Section \ref{sec:experiment} presents the experimental results obtained under both standard and adversarial settings. Section \ref{sec:discussion} provides a detailed discussion and analysis of the observed performance trends, robustness characteristics, and architectural comparisons. Finally, Section \ref{sec:conclusion} concludes the paper and outlines potential directions for future research.

\section{Related Work}
\label{sec:related}

The following sections review three research directions related to our work, namely {\em (i)} static feature-based malware analysis and the EMBER framework,  {\em (ii)} visualization-based representations of malware, and  {\em (iii)}  Mixture of Experts (MoE) and Multi-Gate Mixture of Experts (MMoE) models.

\subsection{EMBER Framework and Static Feature-Based Malware Detection}
To overcome the limitations of traditional signature-based and heuristic approaches that struggle with newly emerging threats, static analysis techniques have been widely adopted. These methods extract features such as byte distributions, entropy measures, and Portable Executable (PE) metadata without executing the file. Feature representations such as EMBER have proven particularly effective in capturing high-dimensional characteristics of executable files \cite{joyce2025ember2024}\cite{ oyama2019ember_features}\cite{gibert2020rise}\cite{csandor2023ember}.

Recent literature extensively leverages static representations to audit classifier boundaries and optimize lightweight feature pipelines. For instance, Gibert et al. \cite{gibert2025assessing} utilize the comprehensive structural feature space of EMBER to benchmark gradient-boosted decision trees against deep learning architectures. Their findings show that metadata-driven representations preserve malware discrimination capabilities even under obfuscation and packing transformations. Similarly, Shinde et al. \cite{shinde2023static} construct a malware detection framework using optimized API import features extracted from the EMBER schema, demonstrating that carefully selected static indicators can achieve high classification accuracy with reduced feature dimensionality.
Recent studies have further demonstrated the effectiveness of EMBER-based high-dimensional static feature representations for adaptive malware classification within a reinforcement learning-driven sequential feature selection framework 1
\cite{khan2026adaptive}.
Recent surveys further highlight that modern malware analysis increasingly requires architectures capable of learning shared yet task-specific representations across multiple cybersecurity objectives \cite{ibrahim2024multitask}\cite{bensaoud2024survey}. Multi-task learning frameworks have shown significant potential in cybersecurity due to their ability to jointly optimize correlated security tasks while improving feature generalization and robustness against evolving attack patterns. In parallel, ensemble and expert-based learning strategies have demonstrated effectiveness in handling heterogeneous feature distributions and reducing model instability in adversarial security environments \cite{dasgupta2022mlcybersecurity}.

Despite these advances, existing malware detection frameworks remain fundamentally constrained by monolithic single-task architectures, where a unified representation is forced to optimize multiple heterogeneous objectives simultaneously. Such approaches often suffer from negative task transfer, particularly when malware family classification, packing detection, and malware versus benign identification depend on distinct structural and semantic characteristics. The Mixture of Experts (MoE) framework proposed in this work addresses this limitation by replacing monolithic processing with expert-specialized learning. More importantly, unlike prior approaches that optimize individual security tasks independently, the proposed framework introduces task-specific gating mechanisms that dynamically route feature representations toward the most relevant experts for each cybersecurity objective.

\subsection{Malware Detection through Visualization-Based Representations}
While static analysis captures high-dimensional features, it remains highly vulnerable to obfuscation techniques like packing and encryption, which drastically alter executable feature distributions \cite{qirui2022packers}. To capture these structural variations without manual engineering, recent advancements rely on vision-based paradigms\cite{brosolo2026security}\cite{brosolo2024sok}. 

Several frameworks map dynamic behavioral data into visual structures. Recent works have further strengthened visual malware analysis through transformer-driven architectures and hybrid spatial learning mechanisms capable of improving representation robustness under complex malware variations \cite{wang2024transmal}.
Concurrently, static binary-to-image mapping approaches avoid runtime emulation constraints but introduce structural scale and data imbalances. Xuan et al. \cite{xuan2024bitcn} combined Opcode, API, and binary sequences into colored RGB images processed via a BiTCN-TAEfficientNet layout, utilizing Atrous Spatial Pyramid Pooling (ASPP) to preserve scale-invariant features. To address class variations, Sharma et al. \cite{sharma2024migan} introduced MIGAN to synthesize minority class samples through a generative pipeline, preserving authentic visual binary markers. 
Meanwhile, Alam et al. \cite{alam2025miracle} developed the \textit{MIRACLE} framework, segmenting binary images into distinct structural regions (headers and sections) and applying an $L_2$ norm CycleGAN to counteract dataset imbalances across varied file sizes. Additional studies have explored lightweight attention-driven CNNs and hierarchical visual encoders to improve malware image representation while reducing computational overhead in large-scale security environments \cite{liu2024attentionmal}\cite{park2025hierarchical}.

Recent methods explore multimodal fusion and decentralized configurations. In \cite{jeon2024static}, the authors use multiple static feature representations and deep learning models to improve malware detection across evolving threat environments. 
Ambekar et al. \cite{ambekar2025fasnet} introduced \textit{FASNet}, a federated adversarial Siamese network configuration across local clients that integrates Fast Gradient Sign Method (FGSM) perturbations to harden image matching under strict data privacy constraints. Emerging studies have also investigated multimodal transformer fusion and distributed malware intelligence frameworks for scalable malware family attribution under adversarial settings \cite{zhang2025multimodal}\cite{kim2024federatedmalware}.

Existing layouts remain constrained by monolithic single-task architectures that process malware representations through static computational pipelines without adaptive expert specialization or conditional routing. A multi-task framework driven by a Mixture of Experts (MoE) configuration remains largely unexplored for malware image analysis, where sparse expert activation could selectively learn discriminative structural patterns across heterogeneous malware representations.

\subsection{Mixture of Experts (MoE) and Multi-Gate Mixture of Experts (MMoE)}

Recent advances in machine learning have significantly improved malware classification by enabling models to learn complex patterns from high-dimensional feature representations. Deep learning approaches, in particular, facilitate automatic extraction of hierarchical features from malware samples. However, conventional single-model architectures often exhibit limited generalization capability across heterogeneous malware distributions, especially in the presence of class imbalance and substantial inter-family variations. Although traditional ensemble techniques improve robustness by combining multiple models, they generally rely on static aggregation strategies and lack adaptive input-dependent specialization. Researchers predominantly design malware classification frameworks using single-model architectures using either static or dynamic feature spaces. But they continue to face challenges in achieving robustness and representation generalization under evolving adversarial malware environments \cite{yan2022survey}. Recent studies have also shown that malware classifiers still struggle to generalize across heterogeneous malware distributions under real-world deployment settings \cite{thirumuruganathan2024detecting}.

Mixture of Experts (MoE) architectures address these limitations through a collection of specialized expert networks coordinated by a gating mechanism that dynamically selects relevant experts for each input sample \cite{jacobs1991moe}\cite{shazeer2017moe}\cite{ fedus2022switch}. This conditional computation strategy enables expert specialization, improves representation learning, and enhances scalability in complex learning environments. Multi-Gate Mixture of Experts (MMoE) architectures further extend this paradigm by introducing task-specific gating networks that independently optimize expert utilization for multiple correlated tasks \cite{ma2018mmoe}. Recent studies have demonstrated the effectiveness of advanced MoE and MMoE frameworks in improving expert routing efficiency, multi-task representation learning, and large-scale model optimization across heterogeneous data distributions \cite{tong2023improved}\cite{wu2024mhmoe}.

Existing malware analysis frameworks predominantly focus on isolated security objectives and monolithic learning strategies, limiting their capability to jointly model multiple interconnected cybersecurity tasks. In malware analysis, tasks such as malware family classification, packed versus unpacked detection, and malware versus benign discrimination often depend on substantially different structural and semantic characteristics. To address this challenge, the proposed framework employs a unified Mixture of Experts architecture that combines expert specialization with adaptive task-aware gating, enabling effective multi-task learning across diverse malware representations.
\section{Proposed Method}
\label{sec:method}

In this work, malware analysis is formulated as a multi-task classification problem involving three interrelated objectives: malware family classification, packed versus unpacked detection, and malware versus benign discrimination. Each task captures a distinct aspect of malware behavior and structure.

Static analysis relies on features extracted from Portable Executable (PE) files, including byte distributions, entropy measures, section characteristics, and imported functions. However, a key challenge arises from the heterogeneous nature of malware data. Packed and unpacked samples exhibit significantly different feature distributions, while malware families often share overlapping characteristics. This leads to difficulties in learning a unified representation using a single model.

Traditional approaches that rely on a single classifier attempt to learn global patterns across all samples, but often fail to generalize effectively on diverse malware distributions. This limitation motivates the need for a model capable of adapting to different data characteristics.

For these reasons, we address this problem by using a Mixture of Experts framework, where multiple specialized expert networks focus on different aspects of the data and a gating mechanism dynamically combines their outputs based on input characteristics.

The proposed framework consists of two main components. First, we investigate two complementary static feature representations extracted from Portable Executable (PE) files, allowing us to evaluate the effectiveness of the proposed approach across different input modalities. Second, we design and compare three Mixture of Experts (MoE)-based architectures that differ in the way expert networks are organized and how task-specific knowledge is shared. Specifically, we analyze a Homogeneous MoE, where all experts share the same architecture; a Heterogeneous MoE, where experts are specialized through different network designs; and a Multi-Gate Mixture of Experts (MMoE), where independent gating mechanisms dynamically route information according to each learning objective. The following subsections describe the adopted feature representations and the proposed expert architectures in detail.

\subsection{Feature Representation}

To effectively characterize executable files without requiring runtime execution, this work employs static feature extraction based on the EMBER framework \cite{csandor2023ember}. EMBER (Elastic Malware Benchmark for Empowering Researchers) provides a comprehensive set of high-dimensional numerical features derived from Portable Executable (PE) files, enabling efficient and scalable malware analysis. For each binary, a 2381-dimensional feature vector is extracted using the \texttt{PEFeatureExtractor} library (version 2), built on top of the LIEF parsing framework. 

The EMBER feature set consists of multiple categories that represent different aspects of a PE file. These include general file metadata, header information, section-level characteristics, imported and exported functions, and byte-level statistical representations. In particular, byte histograms and entropy-based features provide insights into low-level binary distributions, while string-based features capture embedded textual artifacts such as URLs, file paths, and registry keys.

The extracted features can be broadly grouped into the following two categories.
\begin{enumerate}
    \item \textbf{Parsed PE Features:} These features are obtained directly from the PE structure using the LIEF parser. They include file metadata (e.g., size, virtual size), header information, section details, and import/export tables. These features provide high-level structural and semantic information about the executable.
    
    \item \textbf{Format-Agnostic Features:} These features are computed from raw binary data and are independent of the PE format. They include byte histograms, byte-entropy histograms computed using sliding windows, and statistical properties of printable strings. Such features are particularly useful for detecting obfuscation and packing behavior, as they capture irregular patterns in binary distributions.
\end{enumerate}

To ensure numerical stability and consistent feature scaling across all inputs, a two-stage preprocessing pipeline is used. First, a log transformation is applied to the raw feature values to compress highly-tailed distributions and reduce the impact of extreme outliers. Specifically, the transformation is defined as $x^{\prime}=\log(1+x)$, where the constant $1$ is added to ensure that zero-valued features remain defined. Subsequently, min-max normalization is applied to map the transformed values into the $[0, 1]$ range. This combined approach ensures that no single high-magnitude feature dominates the learning process and allows the model to converge more effectively by presenting features on a comparable scale. 

By combining structural, statistical, and behavioral indicators, the EMBER representation provides a rich and comprehensive feature space that is well-suited for training machine learning models in malware classification tasks.

The general composition of the extracted features and their respective categories is summarized in Table~\ref{tab:ember_features}.

\begin{table}
\centering
\scriptsize
\caption{Summary of EMBER Feature Categories}
\label{tab:ember_features}
\begin{tabular}{|l|l|}
\hline
\textbf{Category} & \textbf{Description} \\
\hline
General File Info & File size, metadata, and structural attributes \\
Header Info & Portable Executable header characteristics \\
Section Info & Section names, sizes, and entropy values \\
Imports/Exports & API functions used by the executable \\
Byte Histogram & Frequency distribution of byte values \\
Byte-Entropy & Entropy-based patterns for packing detection \\
String Features & Extracted ASCII strings and statistical properties \\
\hline
\end{tabular}
\end{table}

\begin{figure}[t]
    \centering
    
    \subfloat[\small{Sample from the \texttt{bho} malware family
    (MD5: \texttt{0a60d1a05c0ede7a3ae75df889966d9}).\label{fig:malware_bho}}]{%
        \includegraphics[width=\linewidth]{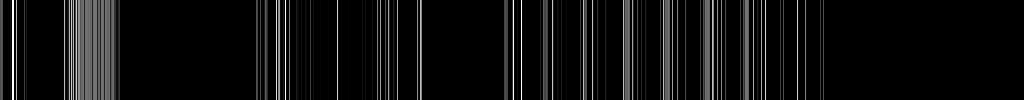}
    }
    
    \vspace{0.3cm}
    
    \subfloat[\small{Sample from the \texttt{vundo} malware family (MD5: \texttt{0a3a5a3bf0b6fedd508e975022753925}).\label{fig:malware_vundo}}]{%
        \includegraphics[width=\linewidth]{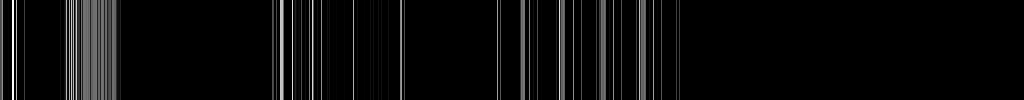}
    }

    \vspace{0.2cm}
    \caption{\small{1D grayscale visualizations generated from the first 1024 bytes of malware samples.}}
    \label{fig:malware_1d_images}
\end{figure}

In addition to structured EMBER features, we evaluate the proposed framework using a 1D image representation (Figure \ref{fig:malware_1d_images}) derived directly from raw executable files. This approach enables the model to learn from low-level binary data without relying on static feature extraction.

The step-by-step procedure for generating this representation is detailed in Algorithm \ref{alg:exe_to_1d}. First, each Portable Executable (PE) file is read as a sequence of raw bytes using the \texttt{readBytes()} operation. Since each byte is represented as an integer in the range [0,255], the executable is naturally converted into a one-dimensional grayscale intensity sequence. The total number of extracted bytes is then computed. Next, the procedure is repeated for predefined sequence lengths of 1024, 4096, and 16384 bytes. If the executable size exceeds the target length $L$, the byte sequence is truncated to retain only the first $L$ bytes. Otherwise, zero-padding is applied to extend shorter sequences to the required fixed length. This ensures uniform input dimensions for deep learning models while preserving as much binary information as possible. Finally, the sequence is reshaped into a $(1,L)$ representation, producing a normalized one-dimensional grayscale image that is directly provided as input to the model.

Raw byte-based representation has been explored in prior work, demonstrating that deep learning models can effectively learn discriminative patterns directly from binary sequences without explicit feature engineering \cite{catanzaro2018malware}\cite{kolosnjaji2016deep}.

\begin{algorithm}[t]
\caption{Generation of a 1D grayscale representation from an executable}
\label{alg:exe_to_1d}

\KwIn{Executable file $f$}
\KwOut{1D grayscale image of size $(1,L)$}

$\mathbf{b}\leftarrow\text{readBytes}(f)$\;

\ForEach{$L\in\{1024,4096,16384\}$}{
    \If{$|\mathbf{b}|>L$}{
        $\mathbf{x}\leftarrow\mathbf{b}[1:L]$\;
    }
    \Else{
        $\mathbf{x}\leftarrow\text{zeroPad}(\mathbf{b},L)$\;
    }
    \Return reshape$(\mathbf{x},(1,L))$\;
}

\end{algorithm}

\subsection{Mixture of Experts Architectures}

In order to appropriately capture the diverse characteristics of malware, we investigate several MoE variants that increasingly improve expert specialization and learning tailored to specific tasks, namely {\em(i)} Homogeneous Mixture of Experts, {\em(ii)} Heterogeneous Mixture of Experts, and {\em(iii)} Multi-Gate Mixture of Experts.

Based on the pioneering research on Adaptive Mixtures of Local Experts \cite{jacobs1991moe}, three different loss formulations were examined to direct the training of distinct experts:

\begin{itemize}
    \item Global Squared Error Loss:
\begin{equation}
E = \left| d - \sum_{i} p_i o_i \right|^2
\label{eq:global_sq}
\end{equation}
where $E$ is the squared error loss\cite{jacobs1991moe}, $d$ is the target output, $p_i$ is the gating weight and $o_i$ is the output of expert $i$. This formulation measures the error between the aggregated expert output and the target $d$ while encouraging strong coupling among experts.

\item Expected Squared Error Loss:
\begin{equation}
E = \sum_{i} p_i \left| d - o_i \right|^2
\label{eq:expected_sq}
\end{equation}
\noindent where all variables are as defined in Equation~\ref{eq:global_sq}, with $E$ here representing the weighted sum of individual expert errors\cite{jacobs1991moe}. This formulation assigns responsibility to each expert based on gating weights, promoting independent learning and specialization.

\item Negative Log-Likelihood (NLL) Loss:
\begin{equation}
E = - \log \left( \sum_{i} p_i \exp\left( - \left| d - o_i \right|^2 \right) \right)
\label{eq:nll}
\end{equation}
where $\exp\left(-|d - o_i|^2\right)$ represents the Gaussian likelihood of expert\cite{jacobs1991moe} $i$ producing the target $d$. This probabilistic formulation models the output as a mixture of Gaussian experts and encourages competitive learning.
\end{itemize}

\subsubsection{Homogeneous Mixture of Experts}
\begin{figure}
\centering
\includegraphics[width=0.5\columnwidth]{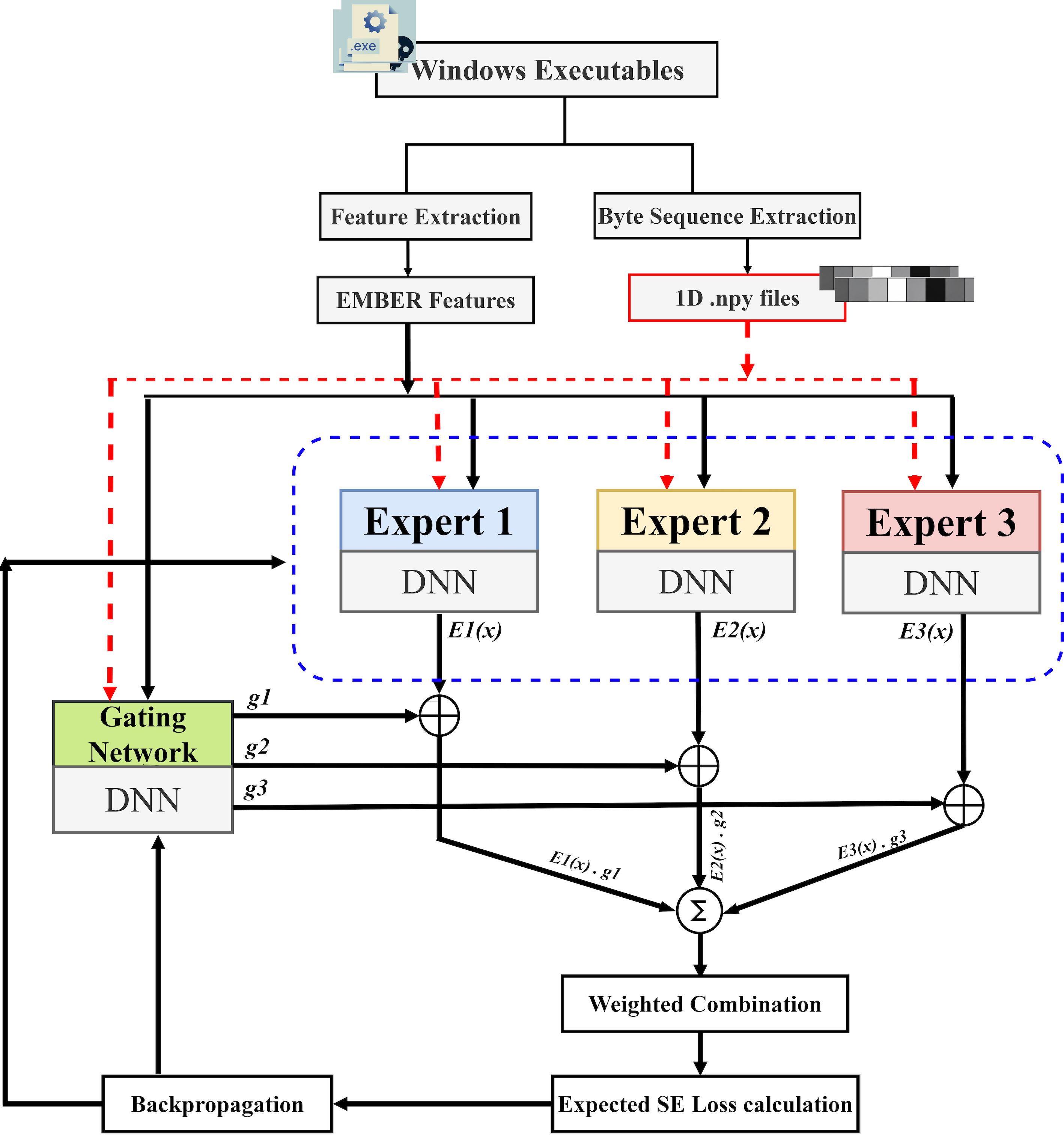}
\caption{The MoE framework for Windows malware classification}
\label{fig:homo_moe}
\end{figure}

The Homogeneous Mixture of Experts (Homo-MoE) model consists of three identical expert networks and a shared gating network, as illustrated in Figure ~\ref{fig:homo_moe}. This figure illustrates the Mixture of Experts (MoE) framework for Windows malware classification using two different input types: {\em(i)} EMBER features and {\em(ii)} 1D image \texttt{.npy} representations extracted from executable files. The EMBER features provide high-level static malware characteristics, while the 1D images preserve low-level binary information. These inputs are processed by three DNN-based expert networks, namely Expert 1, Expert 2, and Expert 3, producing expert outputs $E_1(x)$, $E_2(x)$, and $E_3(x)$, respectively. A gating network dynamically generates routing weights $(g_1, g_2, g_3)$ to determine the contribution of each expert. The weighted outputs $E_1(x)\cdot g_1$, $E_2(x)\cdot g_2$, and $E_3(x)\cdot g_3$ are combined to generate the final prediction. The expected SE loss is computed from the combined output, and backpropagation is performed jointly through both the gating network and the expert networks to optimize the overall MoE framework.  Although the experts share the same architecture, they learn different feature patterns during training, allowing the framework to capture complementary characteristics of malware samples.

The gating network produces a probability distribution over experts:
\begin{equation}
p_i = \frac{\exp(z_i)}{\sum_{j} \exp(z_j)}, 
\end{equation}
\noindent where $p_i$ is the gating weight for expert $i$, $z_i$ is the raw logit score for expert $i$, and $\sum_{j} \exp(z_j)$ is the normalizing term over all experts $j$.

\noindent The final output is computed as:
\begin{equation}
\hat{y} = \sum_{i} p_i \cdot o_i
\end{equation}
\noindent where $\hat{y}$ is the predicted output and all other variables are as defined above.

Among the loss functions assessed, the Expected Squared Error (Equation~\ref{eq:expected_sq}) was selected due to its stability and consistently dependable performance.

To regularize the shared latent representations produced by each expert, a lightweight decoder network is coupled to each expert's 128-dimensional latent output. Each decoder reconstructs the original 2381-dimensional input from the latent space, and a gating-weighted reconstruction loss is computed as:
\begin{equation}
\mathcal{L}_{\text{recon}} = \sum_{i} p_i \cdot \mathbb{E}\left[\left\| x - \hat{x}_i \right\|^2\right]
\end{equation}
\noindent where $x$ is the original input feature vector and $\hat{x}_i$ is the reconstruction produced by the decoder attached to expert $i$. This formulation ensures that each expert's contribution to the reconstruction loss is weighted by its gating responsibility, encouraging the latent space to retain meaningful input-level structure proportional to each expert's activation.

The total training objective combines the task loss with the reconstruction regularizer:
\begin{equation}
\mathcal{L}_{\text{total}} = \mathcal{L}_{\text{mse}} + \alpha \cdot \mathcal{L}_{\text{recon}}
\end{equation}
\noindent where $\alpha$ is a scalar regularization coefficient controlling the contribution of the reconstruction term. 

For the malware versus benign binary classification task, each expert produces a single raw logit output rather than a normalized probability. To convert this into a calibrated confidence score, a sigmoid transformation is applied:
\begin{equation}
P(\text{malware}) = \sigma(z) = \frac{1}{1 + e^{-z}}
\end{equation}
\noindent where $z$ is the raw logit for the malware class. Rather than adopting a fixed decision boundary of 0.5, a threshold sweep over the range $[0.10,\ 0.95]$ in increments of 0.05 is performed, and the threshold that maximizes the macro-averaged F1-score is selected as the operating point. This calibration step adapts the decision boundary to the output distribution of the model, which is particularly important given that logit scales can vary across random seeds and training conditions.

\subsubsection{Heterogeneous Mixture of Experts}
Heterogeneous Mixture of Experts (Hetero-MoE) is a variant of the MoE framework in which multiple experts with different network architectures are combined within a single model. Unlike Homogeneous MoE, where all experts share the same structure, Hetero-MoE employs structurally diverse experts so that each expert can learn different feature representations and decision patterns from the input data. This architectural diversity allows the model to capture complementary information from the feature space more effectively. To improve representation diversity, the Heterogeneous MoE integrates structurally varied expert networks, with each expert designed to address different facets of the feature space.

The Hetero-MoE has the following architecture:
\begin{itemize}
    \item \textbf{Expert 1:} Utilizes skip connections in conjunction with Batch Normalization to improve gradient flow and facilitate the acquisition of deeper representations. This expert produces a 128-dimensional latent representation.
    \item \textbf{Expert 2:} Employs shrinking layer dimensions to aid in the understanding of hierarchical concepts, also producing a 128-dimensional latent representation.
    \item \textbf{Expert 3:} Lowers the input to a representation with fewer dimensions, retaining only the most significant features, and produces a compact 64-dimensional latent representation.
\end{itemize}

\noindent The gating mechanism remains unchanged, where $p_i$ and $\hat{y}$ are as defined in the previous section:
\begin{equation}
p_i = \frac{\exp(z_i)}{\sum_{j} \exp(z_j)}, \qquad \hat{y} = \sum_{i} p_i \cdot o_i
\end{equation}
\noindent In this context, the Negative Log-Likelihood loss (Equation~\ref{eq:nll}) is utilized to promote competitive learning among experts with diverse structures.

The same latent space reconstruction auxiliary objective introduced in Section~III-D is applied to the Heterogeneous MoE. Since the three experts produce latent representations of differing dimensionalities — 128 dimensions for the Residual and Hierarchical experts, and 64 dimensions for the Bottleneck expert — individual decoder networks are constructed for each expert accordingly, mapping each respective latent space back to the original 2381-dimensional input.

The gating-weighted reconstruction loss and the combined training objective follow the same formulation as defined in Section~III-D:
\begin{equation}
\mathcal{L}_{\text{recon}} = \sum_{i} p_i \cdot \mathbb{E}\left[\left\| x - \hat{x}_i \right\|^2\right]
\end{equation}
\begin{equation}
\mathcal{L}_{\text{total}} = \mathcal{L}_{\text{NLL}} + \alpha \cdot \mathcal{L}_{\text{recon}}
\end{equation}
\noindent where $\mathcal{L}_{\text{NLL}}$ is the Negative Log-Likelihood (NLL) loss (Equation~\ref{eq:nll}) and $\alpha$ controls the contribution of the reconstruction term.

\subsubsection{Multi-Gate Mixture of Experts (MMoE)}
The transition from MoE to a Multi-Gate Mixture of Experts (MMoE) is motivated by the inherent complexity of joint optimization in malware analysis. In a multi-task setting, different objectives such as identifying a malware family and detecting packing status often focus on conflicting signals. For instance, while packing detection is heavily influenced by high-level structural anomalies and section entropy, family attribution often requires a deeper analysis of semantic strings and API import patterns.

A fundamental limitation of traditional MoE architectures is the use of a single, unified gating mechanism. By forcing a shared routing strategy across all objectives, the model often suffers from negative transfer, where the gradient updates from one task interfere with the performance of another.

\begin{figure*}
\centering
\includegraphics[width=0.9\textwidth]{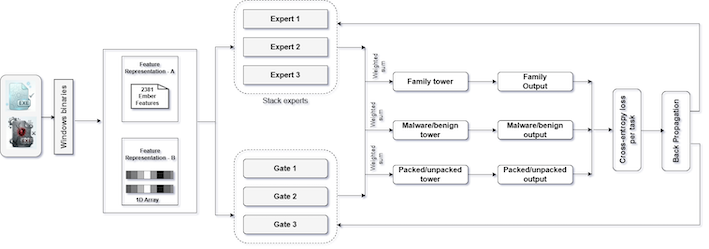} 
\caption{\small{Architecture of the proposed Multi-Gate Mixture of Experts (MMoE) framework for multi-task malware analysis}}
\label{fig:mmoe_architecture1}
\end{figure*}

To mitigate this, we employ the MMoE architecture to decouple the expert routing process visible in Figure \ref{fig:mmoe_architecture1}. Executable files are transformed into feature representations and processed by a pool of shared expert networks that learn complementary malware characteristics. Unlike conventional MoE architectures that rely on a single gating mechanism, MMoE employs independent task-specific gates, allowing each security objective to selectively combine expert outputs according to its requirements. The aggregated representations are passed to dedicated towers for malware family attribution, packed versus unpacked detection, and malware versus benign classification. Task-specific cross-entropy losses are jointly optimized through backpropagation, enabling effective knowledge sharing among experts while mitigating negative transfer between heterogeneous malware analysis tasks. By introducing task-specific gating networks, the framework allows each objective to independently select the most relevant experts for its specific needs. This architectural flexibility enables the model to effectively capture the nuances of each task while still benefiting from the shared knowledge stored in the experts.

The MMoE architecture consists of three key components:
\begin{itemize}
    \item \textbf{Experts:} Shared expert networks that learn general feature representations.
    \item \textbf{Gating Networks:} Task-specific gates that generate expert weight distributions.
    \item \textbf{Towers:} Task-specific neural networks that produce final predictions.
\end{itemize}

The mathematical formulation of the MMoE Architecture is the following. The system processes an input vector $\mathbf{x} \in \mathbb{R}^{D}$ ($D = 2381$) through $E = 3$ shared experts, $T = 3$ tasks, and task-specific prediction towers. First, the input vector $\mathbf{x}$ is passed through each shared expert network $e \in \{1, \dots, E\}$ to extract generic hidden representations:
\begin{equation}
\mathbf{f}_e(\mathbf{x}) = \text{DNN}_e(\mathbf{x})
\label{eq:expert_forward}
\end{equation}
where $\text{DNN}_e$ represents the deep neural network function of the $e$-th expert.

To determine how much each task relies on a given expert, a dedicated gating linear layer outputs raw scores $\mathbf{z}_t$ for each task $t \in \{1, \dots, T\}$. These scores are normalized into a probability routing distribution using a Softmax function:
\begin{equation}
w_{t, e} = \frac{\exp(z_{t, e})}{\sum_{j=1}^{E} \exp(z_{t, j})}
\label{eq:gate_softmax}
\end{equation}
where $z_{t,e}$ is the raw score assigned to expert $e$ by task $t$, and $w_{t, e}$ represents the final calculated gating weight.

Next, a task-specific feature representation $\mathbf{m}_t$ is constructed as a weighted sum of all expert outputs based on the routing parameters from Equation \eqref{eq:gate_softmax}:
\begin{equation}
\mathbf{m}_t = \sum_{e=1}^{E} w_{t, e} \cdot \mathbf{f}_e(\mathbf{x})
\label{eq:mmoe_blend}
\end{equation}
This representation is passed into a prediction tower network to generate final prediction task logits $\hat{\mathbf{y}}_t = \text{DNN}_t^{(\text{tower})}(\mathbf{m}_t)$, where $\text{DNN}_t^{(\text{tower})}$ represents the deep neural network function of the $t$-th task tower.

Each task balances disproportionate class distributions using an inverse-frequency cross-entropy loss $\mathcal{L}_t$:
\begin{equation}
\mathcal{L}_t = -\sum_{c=1}^{C_t} \alpha_{t,c} \cdot y_{t,c} \cdot \log\left(\frac{\exp(\hat{y}_{t,c})}{\sum_{j=1}^{C_t} \exp(\hat{y}_{t,j})}\right)
\label{eq:task_loss}
\end{equation}
where $C_t$ is the total number of classes for task $t$, $\alpha_{t,c}$ is the balance weight factor for class $c$, $y_{t,c}$ is a one-hot class indicator and $\hat{y}_{t,c}$ is the model's predicted logit output for class $c$.

The entire model is optimized simultaneously using a weighted combination of these task-specific loss values, which defines the global objective function $\mathcal{L}_{\text{total}}$:
\begin{equation}
\mathcal{L}_{\text{total}} = \gamma_1 \cdot \mathcal{L}_1 + \gamma_2 \cdot \mathcal{L}_2 + \gamma_3 \cdot \mathcal{L}_3
\label{eq:global_loss}
\end{equation}
where $\gamma_1=1.7$, $\gamma_2=0.3$, and $\gamma_3=1.2$ are empirical task-weighting hyperparameters, and the index mappings $t=1,2,3$ represent the Malware Family, Packed or Unpacked, and Malware or Benign identification tasks, respectively. These specific scaling weights were finalized after extensive empirical experimentation to achieve balanced convergence across all three prediction tasks and prevent any single task from dominating the gradient updates.

During the backward pass, the optimization algorithm calculates errors at the output layer and propagates them backward through the network to update the weights. Let $p_{t,c}$ represent the final predicted probability for class $c$ in task $t$. The initial error signal at the very top of the network-the derivative of the global loss with respect to the raw predicted logits $\hat{y}_{t,c}$ is calculated as:
\begin{equation}
\frac{\partial \mathcal{L}_{\text{total}}}{\partial \hat{y}_{t,c}} = \gamma_t \cdot \alpha_{t,c} \cdot (p_{t,c} - y_{t,c})
\label{eq:grad_logits}
\end{equation}
where $\gamma_t$ corresponds to the specific task weight constants defined in Equation \eqref{eq:global_loss}. This simple subtraction $(p_{t,c} - y_{t,c})$ measures the distance between the prediction and the true label, scaled up or down by our task-weighting and class-balancing factors. Using this error signal, the chain rule of differentiation is applied to determine how much this error affects the task-blended features, resulting in a task-specific error vector denoted as $\frac{\partial \mathcal{L}_{\text{total}}}{\partial \mathbf{m}_t}$.

Because all $E=3$ expert networks are shared, each individual expert affects the performance of all three tasks simultaneously. Therefore, to calculate the total error assigned to a specific expert's output $\mathbf{f}_e(\mathbf{x})$, we must sum the error contributions coming back from all three tasks, multiplied by how much weight that task's gate originally gave to that expert:
\begin{equation}
\frac{\partial \mathcal{L}_{\text{total}}}{\partial \mathbf{f}_e(\mathbf{x})} = \sum_{t=1}^{T} \frac{\partial \mathcal{L}_{\text{total}}}{\partial \mathbf{m}_t} \cdot w_{t,e}
\label{eq:grad_expert_output}
\end{equation}
This error vector combines the feedback from across the entire multi-task network. The optimization algorithm uses this final combined value to update the internal parameters of all the MMoE components, ensuring they adjust their weights to benefit all tasks.

\section{Experimental Campaign}
\label{sec:experiment}

This section presents the performance evaluation of the proposed Mixture of Experts (MoE) models. To assess the framework's effectiveness and robustness, we conducted four distinct experimental series:

\begin{itemize}
    \item {Standard Performance Evaluation:} F1 score, accuracy, and Combined Detection Rate (CDR) are assessed across all MoE variants using original samples to address \textbf{RQ1}.
    \item {Feature Representation Analysis:} The efficacy of structured EMBER features is compared against raw 1D images of varying lengths to determine the optimal input format for \textbf{RQ2}.
    \item {Adversarial Robustness Testing:} Model resilience against metamorphic distribution shifts is evaluated using samples mutated by PyMetaEngine to address the requirements of \textbf{RQ3}.
    \item {MMoE Ablation and Architectural Study:} A systematic evaluation of internal configurations, including shared expert sizes and task-specific tower dimensions, was conducted to identify the optimal hyperparameter settings for addressing \textbf{RQ1}.
\end{itemize}

\subsection{Preliminary Architectural and Regularization Tuning Experiments}
\label{subsec:prelim_experiments}

Prior to executing the standard and adversarial validation stages, a sequence of preliminary empirical experiments was conducted to evaluate model behavior under varying network configurations and loss regularization coefficients. The architectural dimensions and optimization configurations for the EMBER feature and 1D image input type across all evaluated MoE frameworks are detailed in Table~\ref{tab:hyperparameters_ember} and Table~\ref{tab:hyperparameters_1d}, respectively.

\subsubsection{Loss Regularization Coefficient Evaluation ($\alpha$)}
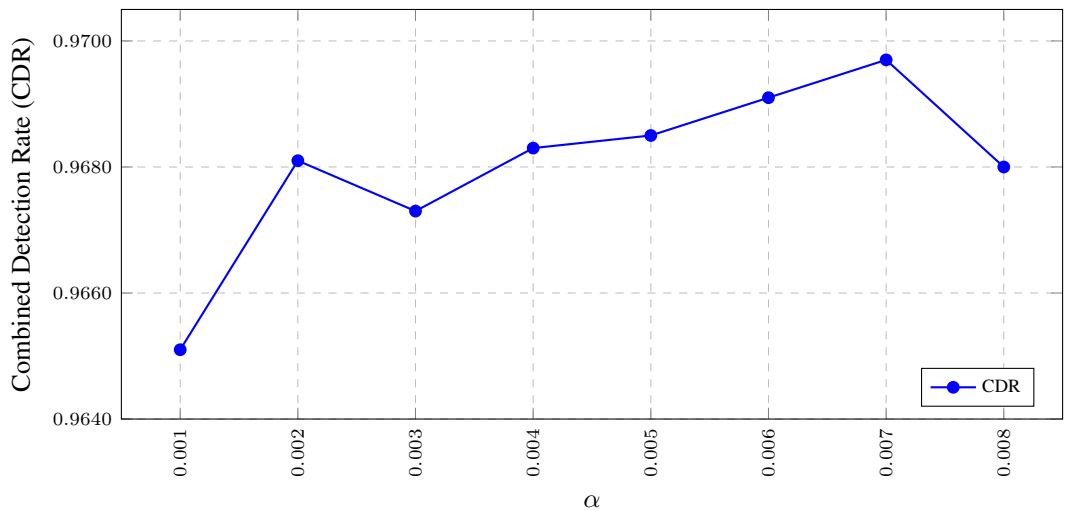
\begin{figure}[ht]
\centering
\begin{tikzpicture}
\begin{axis}[    
    xlabel={$\alpha$},
    ylabel={Combined Detection Rate (CDR)},
    xmin=0.0005, xmax=0.0085,
    ymin=0.9640, ymax=0.9705,
    scaled x ticks=false,
    xtick={0.001, 0.002, 0.003, 0.004, 0.005, 0.006, 0.007, 0.008},
    xticklabel style={
        rotate=90,
        anchor=east,
        font=\scriptsize, 
        /pgf/number format/fixed,
        /pgf/number format/precision=3,
        /pgf/number format/fixed zerofill
    },
    yticklabel style={
        font=\scriptsize, 
        /pgf/number format/fixed,
        /pgf/number format/precision=4,
        /pgf/number format/fixed zerofill
    },
    ymajorgrids=true,
    xmajorgrids=true,
    grid style=dashed,
    width=0.85\linewidth,
    height=7cm,
    legend pos=south east,
    legend style={font=\scriptsize} 
]

\addplot[
    color=blue,
    mark=*,
    mark size=2pt, 
    thick,
    ]
    coordinates {
    (0.001, 0.9651)
    (0.002, 0.9681)
    (0.003, 0.9673)
    (0.004, 0.9683)
    (0.005, 0.9685)
    (0.006, 0.9691)
    (0.007, 0.9697)
    (0.008, 0.9680)
    };
\addlegendentry{CDR}

\end{axis}
\end{tikzpicture}
\caption{Impact of the reconstruction regularization coefficient ($\alpha$) on the Combined Detection Rate (CDR) for the Homogeneous MoE. Higher CDR is achieved at $\alpha = 0.007$.}
\label{graph:alpha_vs_cdr}
\end{figure}

\begin{figure}[ht]
\centering
\begin{tikzpicture}
\begin{axis}[
    xlabel={$\alpha$},
    ylabel={Combined Detection Rate (CDR)},
    xmin=-0.015, xmax=0.22,
    ymin=0.9640, ymax=0.9705,
    scaled x ticks=false,
    xtick={0.001, 0.01, 0.05, 0.08, 0.1, 0.12, 0.15, 0.18},
    xticklabel style={
        rotate=90,
        anchor=east,
        font=\scriptsize,
        /pgf/number format/fixed,
        /pgf/number format/precision=3,
        /pgf/number format/fixed zerofill
    },
    yticklabel style={
        font=\scriptsize,
        /pgf/number format/fixed,
        /pgf/number format/precision=4,
        /pgf/number format/fixed zerofill
    },
    ymajorgrids=true,
    xmajorgrids=true,
    grid style=dashed,
    width=0.85\linewidth,
    height=7cm,
    legend pos=south east,
    legend style={font=\scriptsize}
]
\addplot[
    color=blue,
    mark=*,
    mark size=2pt,
    thick,
]
coordinates {
    (0.001, 0.9674)
    (0.01,  0.9691)
    (0.05,  0.9667)
    (0.08,  0.9681)
    (0.1,   0.9694)
    (0.12,  0.9689)
    (0.15,  0.9684)
    (0.18,  0.9670)
};
\addlegendentry{CDR}
\end{axis}
\end{tikzpicture}
\caption{Impact of the reconstruction regularization coefficient
($\alpha$) on the Combined Detection Rate (CDR) for the Heterogeneous MoE. Higher CDR is achieved at $\alpha = 0.1$.}
\label{graph:alpha_vs_cdr_hetero}
\end{figure}
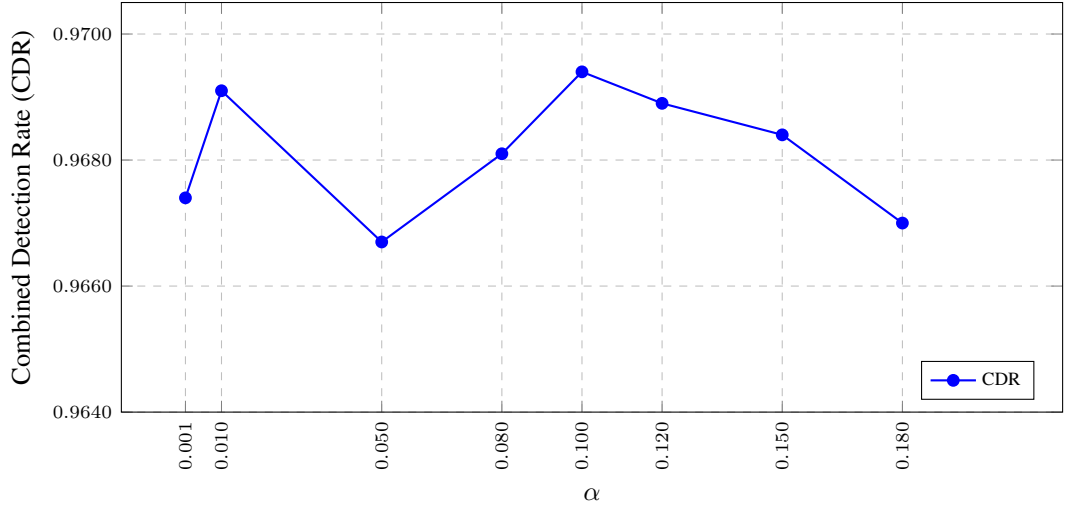

To find the optimal scaling values for the latent space reconstruction objective as explained in Section \ref{sec:method} for Homogeneous and Heterogeneous MoE, a sequence of calibration experiments was performed across different candidate boundaries for $\alpha$ ranging from [0, 1], as illustrated in Fig.~\ref{graph:alpha_vs_cdr}and Fig.~\ref{graph:alpha_vs_cdr_hetero}. This demonstrated that evaluating the system with an exact parameter value of $\alpha = 0.007$ under the Expected Squared Error loss minimized the failure count for Homogeneous MoE, and for Heterogeneous MoE $\alpha = 0.1$ was selected along with the NLL loss while maintaining stable reconstruction across the structurally diverse experts.

\subsubsection{Ablation analysis with different architectural variations in MMoE}
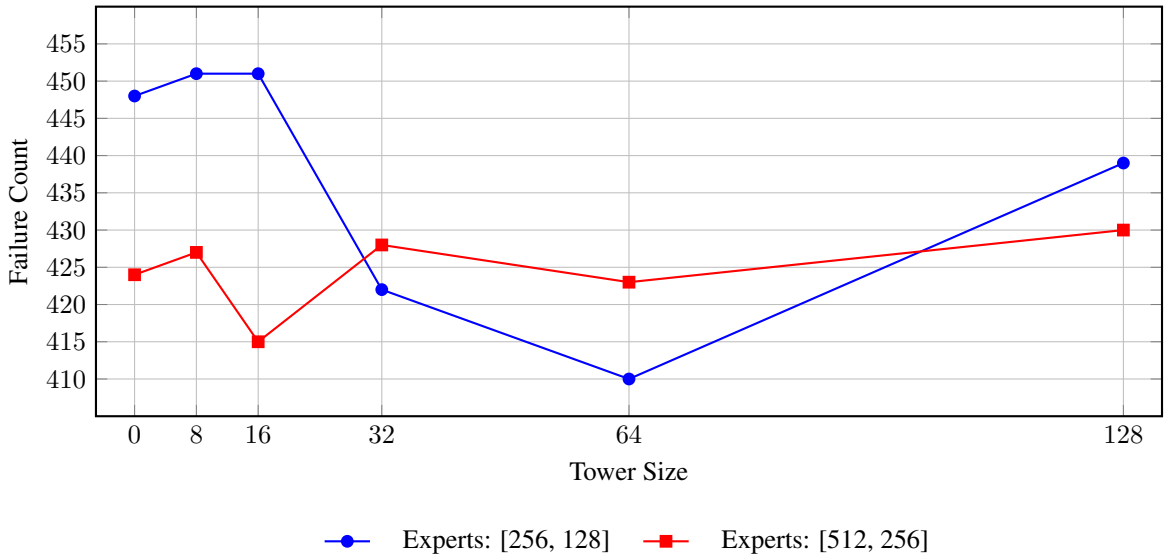
\begin{figure}[htbp]
\centering
    \begin{tikzpicture}
    \begin{axis}[
        width=0.95\columnwidth,
        height=7cm,
        xlabel={Tower Size},
        ylabel={Failure Count},
        xtick={0, 8, 16, 32, 64, 128},
        ytick={410, 415, 420, 425, 430, 435, 440, 445, 450, 455},
        grid=major,
        ymin=405, ymax=460,
        xmin=-5, xmax=133, 
        thick,
        legend style={
            at={(0.5,-0.25)},
            anchor=north,
            legend columns=-1,
            column sep=10pt,
            draw=none
        },
        legend cell align={left}
    ]
    
    \addplot[color=blue, mark=*, thick] coordinates {
        (0,448) (8,451) (16,451) (32,422) (64,410) (128,439)
    };
    \addlegendentry{Experts: [256, 128]}

    \addplot[color=red, mark=square*, thick] coordinates {
        (0,424) (8,427) (16,415) (32,428) (64,423) (128,430)
    };
    \addlegendentry{Experts: [512, 256]}
    
    \end{axis}
    \end{tikzpicture}
\vspace{0.5cm} 
\caption{\small{Ablation analysis of tower size variation across different expert architectures on Dataset D1. The configuration utilizing shared experts of size [256, 128] and tower size of 64 yielded the minimum failure count. }}
\label{fig:hyperparameter_tuning}
\end{figure}

The architectural dimensions for the multi-gate system components were finalized after a comparative analysis involving structural variations of shared expert layers, task-specific gates, and towers. Experiments were run across two distinct baseline expert designs: a compact $[256, 128]$ network and a wider $[512, 256]$ structure. Concurrently, the size parameters for the task-specific towers were scaled across different hidden layer values $\{0, 8, 16, 32, 64, 128\}$ to identify the configuration that minimized multi-task classification errors. The performance curves of the ablation study are plotted in Figure~\ref{fig:hyperparameter_tuning}. The empirical metrics demonstrated that the configuration pairing the $[256, 128]$ shared expert layout with a fixed task tower size of 64 minimized sample classification failures. This layout was selected to balance high-dimensional feature routing accuracy with computational economy.

\subsection{Setup}

The experimental framework was implemented using Python 3.12.7 operating with Ubuntu 22.04 LTS. This specific operating system environment was utilized to ensure compatibility with the EMBER (2018 v2) static feature extraction framework and the LIEF library (v0.14). To generate realistic adversarial scenarios, the PyMetanEgine was used to introduce structural mutations while preserving program semantics. Furthermore, the VirusTotal API was utilized to verify sample classification and maintain the integrity of the ground truth labels across the dataset.

All experiments were conducted using the departmental high-performance computing infrastructure. Model training and evaluation were accelerated using two NVIDIA Tesla A100 Tensor Core GPUs, each equipped with 40 GB of dedicated GPU memory.

\subsection{Dataset Description and Preparation}
\label{subsec:dataset_collection}

The dataset used in this study consists of $66,400$ Portable Executable (PE) files collected from multiple publicly available sources to ensure diversity and realism in malware analysis. Malware samples belonging to 20 different malware families were primarily obtained from MalwareBazaar \cite{malwarebazaar}, which provides continuously updated real-world malware binaries for cybersecurity research. Benign PE files were collected from legitimate software repositories, including PortableApps \cite{portableapps} and the Practical Security Analytics dataset \cite{lester2021dataset}. All benign executables were manually verified and scanned using antivirus engines to confirm their legitimacy and ensure the absence of malicious behavior before inclusion in the dataset.

\begin{figure}
\centering
\includegraphics[width=0.8\linewidth]{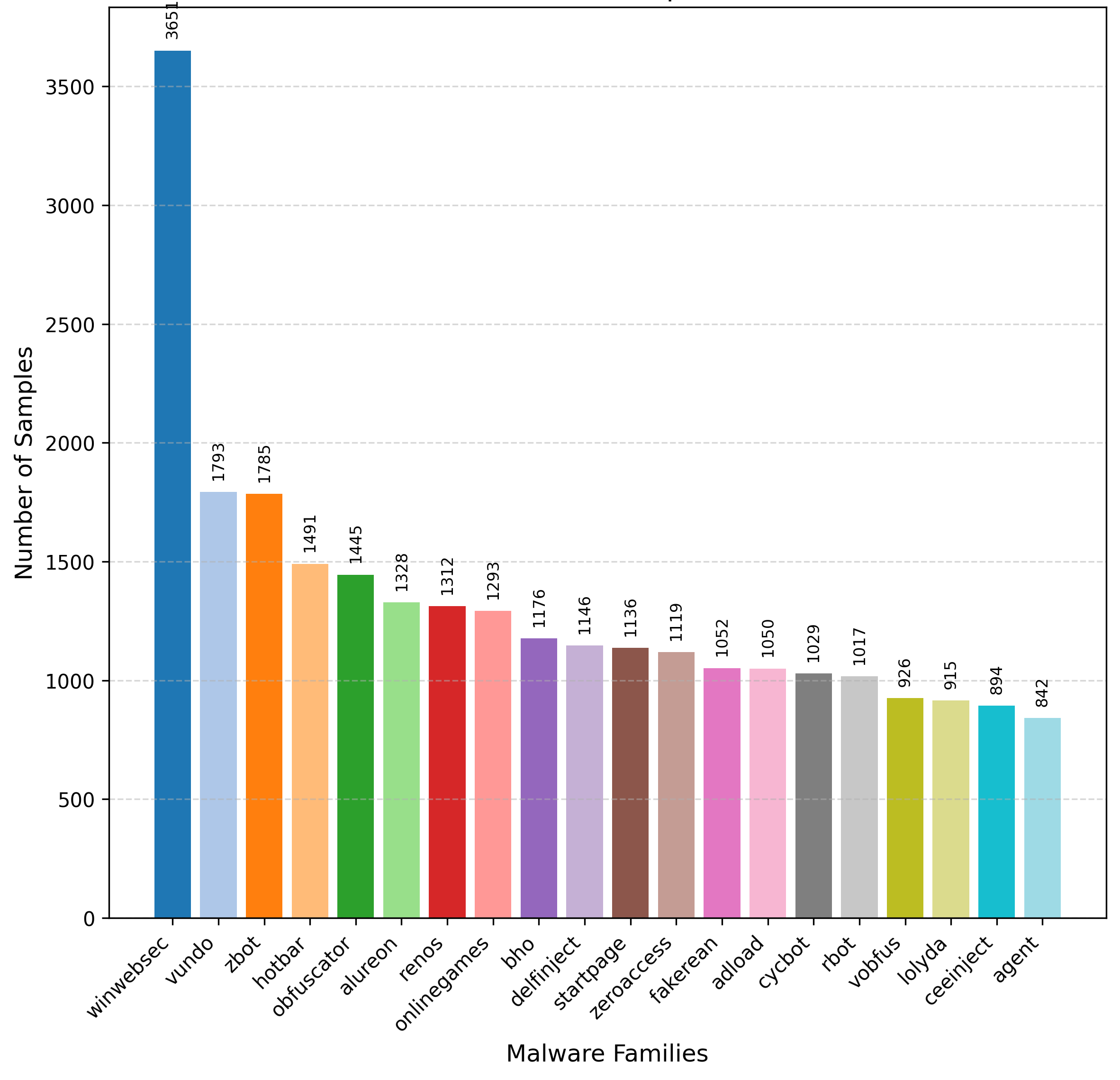}
\caption{\small{Distribution of malware samples across different malware families (excluding benign samples).}}
\label{fig:dataset_distribution}
\end{figure}

The collected dataset contains 26,400 malware samples and 40,000 benign executables, enabling both binary and multi-class malware classification experiments. Before feature extraction, all PE files were validated for structural integrity, and corrupted or unreadable samples were discarded. Feature extraction was performed using the EMBER framework \cite{joyce2025ember2024}, generating a 2381-dimensional feature vector for each executable. The extracted features were normalized using Min-Max scaling to maintain uniform feature distributions during training. The complete dataset was divided into training and testing subsets using an 80:20 split ratio. Figure~\ref{fig:dataset_distribution} illustrates the distribution of malware samples across different families, where noticeable class imbalance can be observed among certain malware categories. This imbalance reflects practical real-world malware distributions and motivates the use of macro-averaged evaluation metrics to ensure fair performance assessment across all classes.

Our dataset is partitioned into standard training and testing subsets, which are referred to as the ``original train'' and ``original test'' sets throughout this paper. To evaluate the proposed models under varied training and evaluation paradigms, five distinct dataset configurations were curated from the primary data repository. The detailed composition of these configurations is summarized in Table~\ref{tab:dataset_configurations}, and the individual datasets are defined as follows:

\begin{itemize}
    \item \textbf{D1:} Comprises the standard EMBER features extracted from $53,120$ original training samples and $13,280$ original test samples to establish a baseline.

    \item \textbf{D2:} We use an augmented training comprising the EMBER features of $53,973$ samples (comprising the 53,120 originals plus 853 augmented samples) and a dedicated adversarial test suite of $1,392$ mutated samples.

    \item \textbf{D3:} Total of $53,973$ training samples of EMBER features, out of which $53,120$ are originals and 853 are mutated samples. The test suite consists of $13,280$ original samples.

    \item \textbf{D4:} The dataset constitutes the first $n$ bytes (where $n \in \{1024, 4096, 16,384\}$) of the binary executables as one-dimensional image, spanning $53,120$ training samples and $13,280$ original test samples.

    \item \textbf{D5:} This dataset maintains the identical 1D byte-level representations as in 'D4' across $53,120$ training instances, but replaces the evaluation suite with $1,392$ mutated, adversarial executable samples.
\end{itemize}

\begin{table*}
\centering
\caption{\small{Dataset configurations used for training and evaluation under standard and adversarial settings.}}
\label{tab:dataset_configurations}
\renewcommand{\arraystretch}{1.2}
\small
\begin{tabular}{c c c c c l}
\toprule
\multirow{2}{*}{\textbf{Dataset}} & \multicolumn{2}{c}{\textbf{Training Samples}} & \multicolumn{2}{c}{\textbf{Test Samples}} & \multirow{2}{*}{\textbf{Representation}} \\
\cmidrule(lr){2-3} \cmidrule(lr){4-5}
 & \textbf{Original} & \textbf{Mutated} & \textbf{Original} & \textbf{Mutated} &  \\
\midrule
D1 & 53,120 & --  & 13,280 & --    & EMBER feature \\
D2 & 53,120 & 853 & --      & 1,392 & EMBER feature  \\
D3 & 53,120 & 853 & 13,280  & --    & EMBER feature  \\
D4 & 53,120 & --  & 13,280  & --    & 1D image  \\
D5 & 53,120 & --  & --      & 1,392 & 1D image  \\
\bottomrule
\end{tabular}
\end{table*}

To analyze the impact of different binary representations on malware classification performance, two feature extraction approaches are considered in this study:

\begin{itemize}
    \item \textbf{EMBER Feature Representation:} A 2381-dimensional structured feature vector extracted using the EMBER framework.
    \item \textbf{1D image Representation:} Raw byte sequences of fixed lengths (1024, 4096, and 16384), derived directly from executable files.
\end{itemize}

The proposed Mixture of Experts models are trained and evaluated separately for each input configuration, allowing a direct comparison between structured feature-based and raw byte-based learning.

\subsection{Experimental Configurations}

To systematically evaluate the performance, adaptability, and adversarial resilience of the proposed MoE frameworks, we conducted five series of experiments, which are listed below.
\begin{itemize}
    \item \textbf{Experiment 1:} We evaluated the standard baseline performance of the models using the EMBER feature vectors from dataset D1.
    \item \textbf{Experiment 2:} We tested the defensive capabilities of the baseline models trained using EMBER feature, against mutated malware samples using dataset D2.
    \item \textbf{Experiment 3:} We measured how training with an expanded EMBER dataset with mutated samples (D3) affects model robustness when tested with original samples.
    \item \textbf{Experiment 4:} We analyzed the baseline performance of the models using raw one-dimensional byte images from dataset D4.
    \item \textbf{Experiment 5:} We evaluated the resilience of the 1D structural models when subjected to adversarial executable samples from dataset D5.
\end{itemize}

\begin{table*}[t]
\centering
\caption{Hyperparameters used across all MoE architectures for the EMBER input representation.}
\label{tab:hyperparameters_ember}

\resizebox{\textwidth}{!}{%
\begin{tabular}{|l|p{4.2cm}|p{7.3cm}|p{4.5cm}|}
\hline
\textbf{Hyperparameter} &
\textbf{Homogeneous MoE} &
\textbf{Heterogeneous MoE} &
\textbf{MMoE} \\
\hline

Number of Experts &
3 &
3 &
3 \\
\hline

Expert Architecture &
Dense(256)--Dense(128)--Latent(128) &
\makecell[l]{
E1: Dense(512)--Dense(512)--Add--Dense(128)\\
E2: Dense(512)--Dense(512)\\--Dense(256)--Dense(256)--Dense(128)\\
E3: Dense(512)--Dense(64)--Dense(512)--Dense(64)
} &
Dense(256)--Dense(128) \\
\hline

Gate Architecture &
\multicolumn{2}{c|}{Dense(64)--BN--ReLU--Softmax(3)} &
\makecell[l]{Dense(64)--Softmax(3)\\one gate per task} \\
\hline

Tower Hidden &
-- &
-- &
Dense(64) per task \\
\hline

L2 Regularization &
$10^{-4}$ &
$10^{-4}$ &
-- \\
\hline

Dropout &
0.3 &
0.3 &
0.2 \\
\hline

Optimizer &
\multicolumn{3}{c|}{Adam} \\
\hline

Learning Rate &
\multicolumn{3}{c|}{$10^{-3}$} \\
\hline

Batch Size &
64 &
64 &
32 \\
\hline

Max Epochs &
100 &
100 &
600 \\
\hline

Early Stopping &
\makecell[l]{Patience = 5\\MinDelta = $10^{-4}$} &
\makecell[l]{Patience = 7\\MinDelta = $10^{-4}$} &
\makecell[l]{Patience = 20\\MinDelta = $10^{-4}$} \\
\hline

Reconstruction Weight ($\alpha$) &
0.007 &
0.1 &
-- \\
\hline

\end{tabular}%
}

\end{table*}

\begin{table*}
\centering
\caption{\small{Hyperparameters used across all MoE architectures for the 1D image input type.}}
\label{tab:hyperparameters_1d}
\resizebox{\textwidth}{!}{%
\begin{tabular}{| l | p{4cm} | p{8cm} | p{5cm} |}
\hline
\textbf{Hyperparameter} & \textbf{Homogeneous MoE} & \textbf{Heterogeneous MoE} & \textbf{MMoE} \\
\hline
Number of Experts       & 3 & 3 & 3 \\
\hline
Expert Architecture     & \makecell[l]{Conv(64,7)--Conv(128,5) \\ --Dense(128)}
                        & \makecell[l]{E1: Conv(32,5)--Conv(64,3)--Dense(128) \\ E2: Conv(64,7)--Conv(128,5)--Conv(128,3)--Dense(256) \\ E3: Conv(16,3)--Dense(64)}
                        & Dense(256--128) \\
\hline
Gate Architecture       & \multicolumn{2}{c|}{Dense(64)--BN--ReLU--Softmax(3)} & \makecell[l]{Dense(64)--Softmax(3), \\ one per task} \\
\hline
Tower Hidden            & -- & -- & Dense(64) per task \\
\hline
L2 Regularization       & $1\times10^{-4}$ & $1\times10^{-4}$ & -- \\
\hline
Dropout                 & 0.3 & 0.3 & 0.2 \\
\hline
Optimizer               & \multicolumn{3}{c|}{Adam} \\
\hline
Learning Rate           & \multicolumn{3}{c|}{$1\times10^{-3}$} \\
\hline
Batch Size              & 32 & 64 & 256 \\
\hline
Max Epochs              & 300 & 150 & 250 \\
\hline
Early Stopping          & {Patience=5, MinDelta=$1\times10^{-4}$} & {Patience=7, MinDelta=$1\times10^{-4}$} & Patience=20, MinDelta=$1\times10^{-4}$ \\
\hline
Reconstruction ($\alpha$) & 0.007 & 0.1 & -- \\
\hline
\end{tabular}%
}
\end{table*}
\subsection{Evaluation Metrics}
\label{subsec:evaluation_metrics}
Model performance is assessed using metrics derived from the confusion matrix, covering both individual task performance and overall multi-task correctness. To account for variability across different random seeds, each experiment is conducted over multiple runs, and both the average ($\mu$) and sample standard deviation ($\sigma$) are reported for all metrics. The average reflects the central tendency of performance, while the standard deviation quantifies the consistency and stability of the model across runs.

\begin{enumerate}
\item \textbf{Accuracy:}  
Accuracy measures the proportion of correctly classified samples among the total number of predictions. In the binary malware detection task, accuracy considers both malware and benign predictions, where True Positives (TP) denote correctly identified malware samples, True Negatives (TN) denote correctly identified benign samples, False Positives (FP) represent benign samples incorrectly classified as malware, and False Negatives (FN) represent malware samples incorrectly classified as benign. The binary classification accuracy is computed as:

\[
\text{Accuracy} = \frac{TP + TN}{TP + TN + FP + FN}
\]

For multi-class tasks such as malware family classification, accuracy represents the proportion of samples assigned to their correct class labels among all evaluated samples. Since malware datasets often exhibit class imbalance, accuracy alone may not fully capture model performance for fine-grained classification tasks. Therefore, additional metrics such as the F1-score are also employed for comprehensive evaluation.

\item \textbf{F1-score:}  
The F1-score is employed for tasks such as malware family classification and packing status detection, where class imbalance may be present. It is computed as the harmonic mean of precision and recall, thereby providing a balanced measure of both false positives and false negatives. Precision represents the proportion of correctly predicted positive samples among all samples predicted as positive, while recall represents the proportion of correctly identified positive samples among all actual positive samples. These are formally defined as:
\[
\text{Precision} = \frac{TP}{TP + FP}
\]

\[
\text{Recall} = \frac{TP}{TP + FN}
\]

The F1-score is then computed as:

\[
\text{F1-score} = \frac{2 \cdot (Precision \cdot Recall)}{Precision + Recall}
\]

\item \textbf{Combined Detection Rate (CDR):}  
Given that the proposed framework performs multi-task analysis — encompassing malware versus benign identification, packing status detection, and malware family classification — a unified evaluation metric termed the \textit{Combined Detection Rate (CDR)} is adopted. CDR measures the proportion of samples for which all task predictions are simultaneously correct. A sample is considered correctly detected only when the model successfully performs all of the following tasks:

\begin{itemize}
    \item \textbf{Task 1:} Malware family classification
    \item \textbf{Task 2:} Packed versus unpacked detection
    \item \textbf{Task 3:} Malware versus benign identification
\end{itemize}

\[
\text{CDR} = \frac{\text{Correct predictions (all tasks)}}{\text{Total samples}}
\]

\item \textbf{Attack Success Rate (ASR):}
The Attack Success Rate measures the proportion of samples for which at least one task is incorrectly predicted by the model.
\[
\text{Attack Success Rate (ASR)} = 1 - \text{CDR}
\]

To ensure reliable and reproducible evaluation, each experimental configuration was conducted using multiple random seed values (42, 123, 500, 750, 1000). All reported metrics are presented as the average performance across these runs, while the corresponding standard deviation is used to measure the consistency and stability of the models under different initialization conditions. Lower standard deviation values indicate more stable learning behavior and improved generalization across repeated experiments.
\end{enumerate}
\subsection{Results}

This section presents the performance analysis of the Homogeneous MoE, Heterogeneous MoE, and Multi-Gate MoE (MMoE) under standard and adversarial conditions, evaluated across multiple random seeds. 

\subsubsection{Performance of EMBER features on Dataset D1}
Evaluating the models under the standard configuration (D1) reveals clear performance differences across the three architectures. The MMoE framework (Table~\ref{tab:mmoe_exp1}) demonstrates the highest baseline performance, achieving a CDR of $0.9670$, yielding better performance than both the Homogeneous MoE ($\text{CDR} = 0.9466$) and the Heterogeneous MoE ($\text{CDR} = 0.9425$), illustrated in Tables~\ref{tab:homo_moe_exp1}) and ~\ref{tab:hetero_moe_exp1} respectively. Consequently, the MMoE minimizes the security risk, showing an ASR that is $1.98\%$ lower than the Homogeneous MoE (which has a $5.33\%$ failure rate) and $2.4\%$ lower than the Heterogeneous MoE ($\text{ASR} = 5.75\%$). 

In terms of task-specific performance, MMoE also performs better than the other two models. For Task 1, MMoE achieves an F1-score of $0.9376$, compared to $0.8842$ for the Homogeneous MoE and $0.8754$ for the Heterogeneous MoE, highlighting its superior capability in distinguishing structurally similar malware families. For Task 2, MMoE, performs with an F1-score of $0.9510$, followed by the Heterogeneous MoE at $0.9233$ and the Homogeneous MoE at $0.9130$. For the Task 3, the three models have competitive performance with F1-scores ranging from $0.9930$ to $0.9998$.
\begin{table}[htbp]
\centering
\small
\framebox[\linewidth]{
\begin{minipage}{0.96\linewidth}
\vspace{2pt}
\textbf{Summary:} MMoE significantly performs better than Homogeneous and Heterogeneous MoE architectures, achieving the highest Combined Detection Rate ($\text{CDR} = 0.9670$), the lowest Attack Success Rate ($\text{ASR} = 3.35\%$), and superior performance across all malware classification tasks.
\vspace{2pt}
\end{minipage}
}
\end{table}

\begin{table*}[t]
\centering
\caption{\small{Results obtained by training the Multi-Gate Mixture of Experts model on the original dataset consisting of 53,120 samples and evaluating it on 13,280 test samples.}}
\label{tab:mmoe_exp1}
\small
\begin{tabular}{c c c c cc cc cc}
\toprule
\multirow{2}{*}{Seed} 
& \multirow{2}{*}{Test time} & \multirow{2}{*}{ASR} & \multirow{2}{*}{CDR}
& \multicolumn{2}{c}{Task 1}
& \multicolumn{2}{c}{Task 2}
& \multicolumn{2}{c}{Task 3} \\
\cmidrule(lr){5-6} \cmidrule(lr){7-8} \cmidrule(lr){9-10}
& (sec) & & & Acc. & F1-score & Acc. & F1-score & Acc. & F1-score \\
\midrule
42   
& 0.0271 & 410 (3.09\%) & 0.9691 & 0.9735 & 0.9382 & 0.9913 & 0.9436 & 0.9950 & 0.9948 \\
123  
& 0.0274 & 479 (3.61\%) & 0.9640 & 0.9723 & 0.9352 & 0.9885 & 0.9396 & 0.9946 & 0.9943 \\
500  
& 0.0272 & 429 (3.23\%) & 0.9676 & 0.9731 & 0.9367 & 0.9898 & 0.9506 & 0.9965 & 0.9963 \\
750  
& 0.0275 & 417 (3.14\%) & 0.9685 & 0.9744 & 0.9395 & 0.9952 & 0.9450 & 0.9902 & 0.9998 \\
1000 
& 0.0273 & 442 (3.33\%) & 0.9667 & 0.9737 & 0.9384 & 0.9892 & 0.9263 & 0.9953 & 0.9950 \\
\midrule
Avg. ($\mu$) 
& 0.0273 & 445 (3.35\%) & 0.9670 & 0.9734 & 0.9376 & 0.9908 & 0.9510 & 0.9943 & 0.9940 \\
Std.Dev. ($\sigma$) 
& $\pm$0.0001 & $\pm$26 (0.19\%) & $\pm$0.0018 & $\pm$0.0008 & $\pm$0.0017 & $\pm$0.0027 & $\pm$0.0261 & $\pm$0.0024 & $\pm$0.0023 \\
\bottomrule
\end{tabular}
\end{table*}

\begin{table*}[t]
\centering
\caption{\small{Performance evaluation of the proposed Homogeneous-MoE framework on the original dataset containing 53,120 training samples and 13,280 test samples under non-adversarial evaluation settings.}}
\label{tab:homo_moe_exp1}
\small
\begin{tabular}{c c c c cc cc cc}
\toprule
\multirow{2}{*}{Seed} 
& \multirow{2}{*}{Test time} & \multirow{2}{*}{ASR} & \multirow{2}{*}{CDR} 
& \multicolumn{2}{c}{Task 1} 
& \multicolumn{2}{c}{Task 2} 
& \multicolumn{2}{c}{Task 3} \\
\cmidrule(lr){5-6} \cmidrule(lr){7-8} \cmidrule(lr){9-10}
& (sec) & & & Acc. & F1-score & Acc. & F1-score & Acc. & F1-score \\
\midrule
42   
& 0.2198 & 678 (5.11\%) & 0.9489 & 0.9593 & 0.8984 & 0.9840 & 0.8840 & 0.9954 & 0.9952 \\
123  
& 0.2094 & 628 (4.73\%) & 0.9527 & 0.9590 & 0.8972 & 0.9890 & 0.9322 & 0.9957 & 0.9955 \\
500  
& 0.2324 & 723 (5.44\%) & 0.9455 & 0.9524 & 0.8814 & 0.9869 & 0.9186 & 0.9948 & 0.9946 \\
750  
& 0.2193 & 598 (4.50\%) & 0.9549 & 0.9612 & 0.9014 & 0.9885 & 0.9270 & 0.9958 & 0.9958 \\
1000 
& 0.1632 & 913 (6.88\%) & 0.9312 & 0.9390 & 0.8427 & 0.9850 & 0.9034 & 0.9948 & 0.9945 \\
\midrule
Avg. ($\mu$) 
& 0.2088 & 708 (5.33\%) & 0.9466 & 0.9542 & 0.8842 & 0.9867 & 0.9130 & 0.9953 & 0.9951 \\
Std.Dev. ($\sigma$) 
& $\pm$0.0244 & $\pm$119 (0.94\%) & $\pm$0.0094 & $\pm$0.0091 & $\pm$0.0245 & $\pm$0.0022 & $\pm$0.0196 & $\pm$0.0004 & $\pm$0.0005 \\
\bottomrule
\end{tabular}
\end{table*}

\begin{table*}[t]
\centering
\caption{\small{Experimental results for the proposed Heterogeneous-MoE framework under a 53,120-sample training and 13,280-original sample testing data split.}}
\label{tab:hetero_moe_exp1}
\small
\begin{tabular}{c c c c cc cc cc}
\toprule
\multirow{2}{*}{Seed} 
& \multirow{2}{*}{Test time} & \multirow{2}{*}{ASR} & \multirow{2}{*}{CDR}
& \multicolumn{2}{c}{Task 1}
& \multicolumn{2}{c}{Task 2}
& \multicolumn{2}{c}{Task 3} \\
\cmidrule(lr){5-6} \cmidrule(lr){7-8} \cmidrule(lr){9-10}
& (sec) & & & Acc. & F1-score & Acc. & F1-score & Acc. & F1-score \\
\midrule
42   
& 0.2412 & 690 (5.20\%) & 0.9480 & 0.9545 & 0.8824 & 0.9881 & 0.9320 & 0.9949 & 0.9947 \\
123  
& 0.2368 & 891 (6.71\%) & 0.9329 & 0.9396 & 0.8528 & 0.9871 & 0.9299 & 0.9942 & 0.9939 \\
500  
& 0.2484 & 655 (4.93\%) & 0.9506 & 0.9551 & 0.8807 & 0.9890 & 0.9274 & 0.9953 & 0.9951 \\
750  
& 0.2391 & 749 (5.64\%) & 0.9435 & 0.9506 & 0.9075 & 0.9863 & 0.9075 & 0.9942 & 0.9939 \\
1000 
& 0.2317 & 831 (6.26\%) & 0.9374 & 0.9435 & 0.8536 & 0.9856 & 0.9195 & 0.9933 & 0.9930 \\
\midrule
Avg. ($\mu$) 
& 0.2394 & 763 (5.75\%) & 0.9425 & 0.9487 & 0.8754 & 0.9872 & 0.9233 & 0.9944 & 0.9941 \\
Std.Dev. ($\sigma$) 
& $\pm$0.0061 & $\pm$98 (0.74\%) & $\pm$0.0073 & $\pm$0.0069 & $\pm$0.0229 & $\pm$0.0014 & $\pm$0.0100 & $\pm$0.0007 & $\pm$0.0005 \\
\bottomrule
\end{tabular}
\end{table*}

\subsubsection{Adversarial Robustness Evaluation of EMBER Features on Dataset D2}

The MMoE framework (Table~\ref{tab:mmoe_adv}) maintains the strongest adversarial defense, achieving a CDR of $0.9342$. This represents a minor drop from its standard baseline performance ($\text{CDR} = 0.9670$ on D1). In comparison, the Homogeneous MoE (Table~\ref{tab:homo_adv}) and Heterogeneous MoE (Table~\ref{tab:hetero_adv}) experience much sharper declines, with their CDRs falling to $0.8635$ and $0.8577$, respectively. Consequently, MMoE limits the Attack Success Rate (ASR) to $6.58\%$, which is lower than that of the Homogeneous MoE ($\text{ASR} = 13.66\%$) and the Heterogeneous MoE ($\text{ASR} = 14.23\%$). Heterogeneous MoE also exhibits the highest standard deviation across experimental seeds, indicating reduced stability when processing adversarial inputs.

A task-specific comparison reveals that mutation-based transformations affect the complex malware family classification more severely than the other tasks, yet MMoE handles this distribution shift most effectively. For Task 1, MMoE achieves an F1-score of $0.8985$, whereas the Homogeneous MoE drops to $0.8269$ and the Heterogeneous MoE to $0.8245$. For Task 2, MMoE sustains an F1-score of $0.9504$, while the Homogeneous and Heterogeneous models present F1-scores of $0.9426$ and $0.9293$, respectively. For Task 3, all three frameworks continue to exhibit stable and competitive performance as in the baseline experiments conducted using dataset D1. 
\begin{table}[ht]
\centering
\small
\framebox[\linewidth]{
\begin{minipage}{0.96\linewidth}
\vspace{2pt}
\textbf{Summary:} Under adversarial mutations (D2), MMoE demonstrates resilience, maintaining a high $\text{CDR} = 0.9342$ and limiting the security risk ($\text{ASR} = 6.58\%$), whereas Homogeneous and Heterogeneous MoE models suffer severe degradation with ASRs up to $13\%$ along with reduced stability.
\vspace{2pt}
\end{minipage}
}
\end{table}   

\begin{table*}[t]
\centering
\caption{\small{Performance evaluation of the proposed Multi-Gate Mixture of Experts (MMoE) framework under adversarial attack settings on 1,392 mutated malware samples across multiple random seed initializations.}}
\label{tab:mmoe_adv}
\small
\begin{tabular}{c c c c c cc cc c}
\toprule
\multirow{2}{*}{Seed} 
& \multirow{2}{*}{Test time} & \multirow{2}{*}{ASR} & \multirow{2}{*}{CDR}
& \multicolumn{2}{c}{Task 1}
& \multicolumn{2}{c}{Task 2}
& \multicolumn{1}{c}{Task 3} \\
\cmidrule(lr){5-6} \cmidrule(lr){7-8} \cmidrule(lr){9-9}
& (sec) & & & Acc. & F1-score & Acc. & F1-score & Acc. \\
\midrule
42    
& 0.0274 & 94 (6.76\%) & 0.9324 & 0.9483 & 0.9012 & 0.9770 & 0.9335 & 0.9987 \\
123   
& 0.0271 & 98 (7.05\%) & 0.9295 & 0.9440 & 0.8953 & 0.9820 & 0.9541 & 0.9974 \\
500   
& 0.0275 & 92 (6.61\%) & 0.9339 & 0.9447 & 0.8931 & 0.9806 & 0.9607 & 0.9998 \\
750   
& 0.0272 & 94 (6.75\%) & 0.9325 & 0.9483 & 0.8976 & 0.9820 & 0.9568 & 0.9968 \\
1000  
& 0.0273 & 80 (5.75\%) & 0.9425 & 0.9540 & 0.9054 & 0.9820 & 0.9471 & 0.9988 \\
\midrule
Avg. ($\mu$) 
& 0.0273 & 92 (6.58\%) & 0.9342 & 0.9479 & 0.8985 & 0.9807 & 0.9504 & 0.9983 \\
Std.Dev. ($\sigma$) 
& $\pm$0.0001 & $\pm$7 (0.49\%) & $\pm$0.0049 & $\pm$0.0040 & $\pm$0.0049 & $\pm$0.0022 & $\pm$0.0107 & $\pm$0.0012 \\
\bottomrule
\end{tabular}
\end{table*}

\begin{table*}[t]
\centering
\caption{\small{Adversarial robustness evaluation of the proposed Homogeneous Mixture of Experts (HOMO-MoE) framework on 1,392 mutated malware samples.}}
\label{tab:homo_adv}
\small
\begin{tabular}{c c c c c cc cc c}
\toprule
\multirow{2}{*}{Seed} 
& \multirow{2}{*}{Test time} & \multirow{2}{*}{ASR} & \multirow{2}{*}{CDR}
& \multicolumn{2}{c}{Task 1}
& \multicolumn{2}{c}{Task 2}
& \multicolumn{1}{c}{Task 3} \\
\cmidrule(lr){5-6} \cmidrule(lr){7-8} \cmidrule(lr){9-9}
& (sec) & & & Acc. & F1-score & Acc. & F1-score & Acc. \\
\midrule
42   
& 0.0712 & 148 (10.63\%) & 0.8937 & 0.9167 & 0.8639 & 0.9713 & 0.9473 & 0.9961 \\
123  
& 0.0655 & 177 (12.73\%) & 0.8728 & 0.8836 & 0.8279 & 0.9784 & 0.9577 & 0.9976 \\
500  
& 0.0758 & 187 (13.43\%) & 0.8656 & 0.8857 & 0.8709 & 0.9755 & 0.9448 & 0.9923 \\
750  
& 0.0834 & 219 (15.75\%) & 0.8426 & 0.8563 & 0.7860 & 0.9784 & 0.9317 & 0.9955 \\
1000 
& 0.0834 & 219 (15.75\%) & 0.8426 & 0.8563 & 0.7860 & 0.9784 & 0.9317 & 0.9976 \\
\midrule
Avg. ($\mu$) 
& 0.0759 & 190 (13.66\%) & 0.8635 & 0.8797 & 0.8269 & 0.9764 & 0.9426 & 0.9958 \\
Std.Dev. ($\sigma$) 
& $\pm$0.0073 & $\pm$30 (2.17\%) & $\pm$0.0217 & $\pm$0.0251 & $\pm$0.0408 & $\pm$0.0031 & $\pm$0.0111 & $\pm$0.0021 \\
\bottomrule
\end{tabular}
\end{table*}

\begin{table*}[t]
\centering
\caption{\small{Heterogeneous MoE -- Results obtained on Adversarial Attack with mutated 1392 malware samples. }}
\label{tab:hetero_adv}
\small
\begin{tabular}{c c c c c cc cc c}
\toprule
\multirow{2}{*}{Seed} 
& \multirow{2}{*}{Test time} & \multirow{2}{*}{ASR} & \multirow{2}{*}{CDR}
& \multicolumn{2}{c}{Task 1}
& \multicolumn{2}{c}{Task 2}
& \multicolumn{1}{c}{Task 3} \\
\cmidrule(lr){5-6} \cmidrule(lr){7-8} \cmidrule(lr){9-9}
& (sec) & & & Acc. & F1-score & Acc. & F1-score & Acc. \\
\midrule
42   
& 0.0655 & 188 (13.51\%) & 0.8649 & 0.8735 & 0.8148 & 0.9791 & 0.9539 & 0.9988 \\
123  
& 0.0677 & 206 (14.80\%) & 0.8520 & 0.8649 & 0.8062 & 0.9748 & 0.9096 & 0.9912 \\
500  
& 0.0738 & 288 (20.69\%) & 0.7931 & 0.8146 & 0.7952 & 0.9669 & 0.9054 & 0.9942 \\
750  
& 0.0689 & 146 (10.49\%) & 0.8951 & 0.9130 & 0.8523 & 0.9748 & 0.9485 & 0.9961 \\
1000 
& 0.0799 & 162 (11.64\%) & 0.8836 & 0.9123 & 0.8540 & 0.9647 & 0.9293 & 0.9932 \\
\midrule
Avg. ($\mu$) 
& 0.0711 & 198 (14.23\%) & 0.8577 & 0.8757 & 0.8245 & 0.9721 & 0.9293 & 0.9947 \\
Std.Dev. ($\sigma$) 
& $\pm$0 & $\pm$55 (3.98\%) & $\pm$0.0398 & $\pm$0.0406 & $\pm$0.0271 & $\pm$0.0060 & $\pm$0.0220 & $\pm$0.0029 \\
\bottomrule
\end{tabular}
\end{table*}
\begin{table*}
\centering
\caption{Comprehensive ablation analysis of mutation-based data augmentation scaling across multi-task classification metrics (Test Size $N = 1,392$, Base Training Size = $53,120$).}
\label{tab:augmentation_results}
\small
\begin{tabular}{|l|r|r|r|r|r|r|r|}
\hline
\textbf{Metric / Dataset Composition} & \textbf{3\%} & \textbf{4\%} & \textbf{5\%} & \textbf{6\%} & \textbf{7\%} & \textbf{10\%} & \textbf{20\%} \\
\hline
Mutated samples added    & 644      & 853      & 1,065    & 1,276    & 1,487    & 2,122    & 4,232    \\
Final train size         & 53,764   & 53,973   & 54,185   & 54,396   & 54,607   & 55,242   & 57,352   \\
\hline
\multicolumn{8}{|c|}{\textbf{MoE Task Metrics}} \\
\hline
Family Accuracy          & 0.8872   & \textbf{0.9167}   & 0.8872   & 0.7234   & 0.8161   & 0.6803   & 0.3341   \\
Family F1 Macro          & 0.8675   & 0.8639   & 0.8276   & 0.7012   & 0.7448   & 0.6758   & 0.3004   \\
Family F1 Weighted       & 0.8813   & 0.9185   & 0.8860   & 0.7201   & 0.8130   & 0.6845   & 0.3083   \\
Packer Accuracy          & 0.9763   & 0.9713   & 0.9497   & 0.9626   & 0.9784   & 0.9655   & 0.9203   \\
Packer F1 Macro          & 0.9403   & 0.9473   & 0.8916   & 0.9260   & 0.9590   & 0.9187   & 0.6778   \\
Packer F1 Weighted       & 0.9762   & 0.9709   & 0.9481   & 0.9621   & 0.9784   & 0.9648   & 0.9104   \\
Label Accuracy           & 1.0000   & 1.0000   & 1.0000   & 1.0000   & 1.0000   & 1.0000   & 1.0000   \\
\hline
\multicolumn{8}{|c|}{\textbf{Overall Performance}} \\
\hline
Detection Rate           & 87.28\%  & \textbf{89.37\%}  & 85.20\%  & 70.26\%  & 80.89\%  & 66.59\%  & 28.59\%  \\
Fully Correct            & 1,215    & \textbf{1,244}    & 1,186    & 978      & 1,126    & 927      & 398      \\
Failed Samples           & 177      & \textbf{148}      & 206      & 414      & 266      & 465      & 994      \\
\hline
\end{tabular}
\end{table*}
\subsubsection{Impact of Mutation-Based Data Augmentation on Dataset D3 using EMBER Features}

For generating the mutation-based dataset D3, a particular $4\%$ mutation rate was chosen based on the performance metrics shown in Table ~\ref{tab:augmentation_results}. In models that are trained on this augmented train set, the Multi-gate Mixture-of-Experts (MMoE) framework demonstrates higher optimization compared to the experiments conducted on the original dataset D1 across every model. Here, the MMoE framework (Table~\ref{tab:mmoe_exp2}) resulted in a CDR of $0.9744$. In comparison, the Homogeneous MoE (Table~\ref{tab:homo_moe_exp2}) and Heterogeneous MoE (Table~\ref{tab:hetero_moe_exp2}) have not shown any increments, with their CDRs only reaching up to $0.9343$ and $0.9171$, which is lower than the results obtained in the original dataset D1. Thereby, MMoE limits the Attack Success Rate (ASR) to $2.56\%$, outperforming the Homogeneous MoE ($\text{ASR} = 6.57\%$) and the Heterogeneous MoE ($\text{ASR} = 8.29\%$). Also, when we analyze the comparison among tasks, the addition of mutated samples improved the classification only in MMoE, as MMoE obtains an F1-score of $0.9505$ for Task 1, whereas the Homogeneous MoE reduced to $0.8554$ and the Heterogeneous MoE to $0.8070$. For Task 2, MMoE leads with an average F1-score of $0.9704$, followed by the Homogeneous MoE at $0.9027$ and the Heterogeneous MoE at $0.8982$. For Task 3, all three architectures maintain robust and convergent performance, as they exhibit F1-scores ranging from $0.9941$ to $0.9959$.
\begin{table}[ht]
\centering
\small
\framebox[\linewidth]{
\begin{minipage}{0.96\linewidth}
\vspace{2pt}
\textbf{Summary:} Data augmentation uniquely benefits the MMoE framework, optimizing its performance to a peak $\text{CDR} = 0.9744$ and a minimal $\text{ASR} = 2.56\%$, whereas Homogeneous and Heterogeneous MoE architectures fail to leverage the augmented data, resulting in performance degradation.
\vspace{2pt}
\end{minipage}
}
\end{table}

\begin{table*}[t]
\centering
\caption{\small{Multi-Gate MoE results after mixing 4\% of the mutated samples into the training set and testing on 13,280 samples.}}
\label{tab:mmoe_exp2}
\small
\begin{tabular}{c c c c cc cc cc}
\toprule
\multirow{2}{*}{Seed} 
& \multirow{2}{*}{Test time} & \multirow{2}{*}{ASR} & \multirow{2}{*}{CDR}
& \multicolumn{2}{c}{Task 1}
& \multicolumn{2}{c}{Task 2}
& \multicolumn{2}{c}{Task 3} \\
\cmidrule(lr){5-6} \cmidrule(lr){7-8} \cmidrule(lr){9-10}
& (sec) & & & Acc. & F1-score & Acc. & F1-score & Acc. & F1-score \\
\midrule
42   
& 0.0271 & 345 (2.60\%) & 0.9740 & 0.9795 & 0.9508 & 0.9915 & 0.9455 & 0.9972 & 0.9968 \\
123  
& 0.0277 & 349 (2.63\%) & 0.9737 & 0.9789 & 0.9481 & 0.9916 & 0.9583 & 0.9969 & 0.9967 \\
500  
& 0.0271 & 353 (2.66\%) & 0.9734 & 0.9787 & 0.9493 & 0.9922 & 0.9619 & 0.9969 & 0.9966 \\
750  
& 0.0272 & 325 (2.45\%) & 0.9755 & 0.9809 & 0.9539 & 0.9923 & 0.9503 & 0.9971 & 0.9969 \\
1000 
& 0.0271 & 327 (2.47\%) & 0.9753 & 0.9794 & 0.9503 & 0.9973 & 0.9972 & 0.9927 & 0.9925 \\
\midrule
Avg. ($\mu$) 
& 0.0272 & 339 (2.56\%) & 0.9744 & 0.9795 & 0.9505 & 0.9930 & 0.9704 & 0.9962 & 0.9959 \\
Std.Dev. ($\sigma$) 
& $\pm$0.0002 & $\pm$13 (0.10\%) & $\pm$0.0010 & $\pm$0.0009 & $\pm$0.0022 & $\pm$0.0024 & $\pm$0.0251 & $\pm$0.0019 & $\pm$0.0016 \\
\bottomrule
\end{tabular}
\end{table*}

\begin{table*}[t]
\centering
\caption{\small{Evaluation of the proposed Homogeneous-MoE framework trained on 53,120 original samples with 4\% mutation-based augmentation and tested on 13,280 original samples. }}
\label{tab:homo_moe_exp2}
\small
\begin{tabular}{c c c c cc cc cc}
\toprule
\multirow{2}{*}{Seed} 
& \multirow{2}{*}{Test time} & \multirow{2}{*}{ASR} & \multirow{2}{*}{CDR}
& \multicolumn{2}{c}{Task 1}
& \multicolumn{2}{c}{Task 2}
& \multicolumn{2}{c}{Task 3} \\
\cmidrule(lr){5-6} \cmidrule(lr){7-8} \cmidrule(lr){9-10}
& (sec) & & & Acc. & F1-score & Acc. & F1-score & Acc. & F1-score \\
\midrule
42   
& 0.2760 & 662 (4.98\%)  & 0.9502 & 0.9647 & 0.9123 & 0.9790 & 0.8732 & 0.9961 & 0.9959 \\
123  
& 0.5615 & 940 (7.08\%)  & 0.9292 & 0.9354 & 0.8458 & 0.9875 & 0.9397 & 0.9950 & 0.9948 \\
500  
& 0.5563 & 1038 (7.82\%) & 0.9218 & 0.9345 & 0.8221 & 0.9806 & 0.8419 & 0.9955 & 0.9951 \\
750  
& 0.5727 & 807 (6.08\%)  & 0.9392 & 0.9445 & 0.8534 & 0.9902 & 0.9398 & 0.9962 & 0.9957 \\
1000 
& 0.2970 & 915 (6.89\%)  & 0.9311 & 0.9404 & 0.8436 & 0.9847 & 0.9188 & 0.9938 & 0.9936 \\
\midrule
Avg. ($\mu$) 
& 0.4527 & 872 (6.57\%) & 0.9343 & 0.9439 & 0.8554 & 0.9844 & 0.9027 & 0.9953 & 0.9949 \\
Std.Dev. ($\sigma$) 
& $\pm$0.1448 & $\pm$143 (1.08\%) & $\pm$0.0108 & $\pm$0.0123 & $\pm$0.0338 & $\pm$0.0047 & $\pm$0.0435 & $\pm$0.0009 & $\pm$0.0008 \\
\bottomrule
\end{tabular}
\end{table*}

\begin{table*}[t]
\centering
\caption{\small{Results for the proposed HETERO-MoE framework, utilizing a training set of 53,120 original samples supplemented with 4\% mutation-based data and a validation set of 13,280 original test samples.}}
\label{tab:hetero_moe_exp2}
\small
\begin{tabular}{c c c c cc cc cc}
\toprule
\multirow{2}{*}{Seed} 
& \multirow{2}{*}{Test time} & \multirow{2}{*}{ASR} & \multirow{2}{*}{CDR}
& \multicolumn{2}{c}{Task 1}
& \multicolumn{2}{c}{Task 2}
& \multicolumn{2}{c}{Task 3} \\
\cmidrule(lr){5-6} \cmidrule(lr){7-8} \cmidrule(lr){9-10}
& (sec) & & & Acc. & F1-score & Acc. & F1-score & Acc. & F1-score \\
\midrule
42   
& 0.5292 & 1179 (8.88\%)  & 0.9112 & 0.9185 & 0.7966 & 0.9869 & 0.9118 & 0.9960 & 0.9957 \\
123  
& 0.5292 & 1637 (12.33\%) & 0.8767 & 0.8837 & 0.6908 & 0.9848 & 0.8423 & 0.9951 & 0.9948 \\
500  
& 0.5292 & 1527 (11.50\%) & 0.8850 & 0.8985 & 0.7510 & 0.9643 & 0.8897 & 0.9904 & 0.9901 \\
750  
& 0.5292 & 620 (4.67\%)   & 0.9533 & 0.9599 & 0.8951 & 0.9896 & 0.9301 & 0.9951 & 0.9950 \\
1000 
& 0.5292 & 542 (4.08\%)   & 0.9592 & 0.9622 & 0.9013 & 0.9810 & 0.9172 & 0.9957 & 0.9953 \\
\midrule
Avg. ($\mu$) 
& 0.5292 & 1101 (8.29\%) & 0.9171 & 0.9246 & 0.8070 & 0.9813 & 0.8982 & 0.9945 & 0.9941 \\
Std.Dev. ($\sigma$) 
& $\pm$0 & $\pm$505 (3.80\%) & $\pm$0.0380 & $\pm$0.0355 & $\pm$0.0914 & $\pm$0.0100 & $\pm$0.0345 & $\pm$0.0023 & $\pm$0.0020 \\
\bottomrule
\end{tabular}
\end{table*}


\subsubsection{Performance of Raw 1D image Representation on Dataset D4}
Dataset D4 utilized raw 1D byte-level representations extracted directly from the binary executables. Here, the Heterogeneous model(Table~\ref{tab:hetero_1d_avg}) maintained an initial detection capability at a byte length of 1024, achieving a CDR of $0.9585$, followed by Homogeneous MoE ($\text{CDR} = 0.9576$, Table~\ref{tab:homo_1d_avg}) and the Multi-Gate MoE ($\text{CDR} = 0.9478$, Table~\ref{tab:mmoe_1d_avg}).
Consequently, the Heterogeneous MoE reduces the Attack Success Rate (ASR) to $4.14\%$, which is lower than the Homogeneous MoE ($\text{ASR} = 4.23\%$) and the MMoE framework ($\text{ASR} = 5.22\%$). 

In the case of task-wise comparison, the Heterogeneous MoE achieves an average F1-score of $0.9361$, whereas in the Homogeneous MoE and MMoE framework, it averaged to $0.9272$ and $0.9205$ for Task 1. While for Task 2, MMoE sustains a superior F1-score of $0.9641$, the Homogeneous and Heterogeneous models present lower F1-scores of $0.9020$ and $0.8799$. For Task 3, both the Homogeneous and Heterogeneous architectures maintain robust and convergent performance, as they exhibit F1-scores ranging from $0.9899$ to $0.9921$, while MMoE encounters a more noticeable reduction to $0.9075$.

\begin{table}[ht]
\centering
\small
\framebox[\linewidth]{
\begin{minipage}{0.96\linewidth}
\vspace{2pt}
\textbf{Summary:} The transition to raw 1D images at a baseline length of 1024 advances both the Heterogeneous and Homogeneous MoE models, optimizing their performance to peak $\text{CDR}$ values of $0.9585$ and $0.9576$ alongside minimal failure rates of $4.14\%$ and $4.23\%$ respectively, whereas the MMoE framework struggles to stabilize across all tasks within this raw sequential paradigm.
\vspace{2pt}
\end{minipage}
}
\end{table}

\begin{table*}[t]
\centering
\caption{\small{Heterogeneous MoE Comparison across Byte Lengths (Average of Five Seeds).}}
\label{tab:hetero_1d_avg}
\begin{tabular}{c c c c cc cc cc}
\toprule
\multirow{2}{*}{Byte Length} 
& \multirow{2}{*}{Test time} & \multirow{2}{*}{ASR} & \multirow{2}{*}{CDR}
& \multicolumn{2}{c}{Task 1}
& \multicolumn{2}{c}{Task 2}
& \multicolumn{2}{c}{Task 3} \\
\cmidrule(lr){5-6} \cmidrule(lr){7-8} \cmidrule(lr){9-10}
& (sec) & & & Acc. & F1 & Acc. & F1 & Acc. & F1 \\
\midrule
1024  
& 154.13 & 550 (4.14\%) & 0.9585 & 0.9703 & 0.9361 & 0.9798 & 0.8799 & 0.9903 & 0.9899 \\
4096  
& 224.21 & 853 (6.42\%) & 0.9357 & 0.9501 & 0.8948 & 0.9693 & 0.8759 & 0.9827 & 0.9819 \\
16384 
& 250.95 & 778 (5.85\%) & 0.9413 & 0.9547 & 0.9043 & 0.9732 & 0.8888 & 0.9839 & 0.9832 \\
\bottomrule
\end{tabular}
\end{table*}

\begin{table*}[t]
\centering
\caption{\small{Homogeneous MoE Evaluation across Byte Lengths (Average of Five Seeds).}}
\label{tab:homo_1d_avg}
\begin{tabular}{c c c c cc cc cc}
\toprule
\multirow{2}{*}{Byte Length} 
& \multirow{2}{*}{Test time} & \multirow{2}{*}{ASR} & \multirow{2}{*}{CDR}
& \multicolumn{2}{c}{Task 1}
& \multicolumn{2}{c}{Task 2}
& \multicolumn{2}{c}{Task 3} \\
\cmidrule(lr){5-6} \cmidrule(lr){7-8} \cmidrule(lr){9-10}
& (sec) & & & Acc. & F1 & Acc. & F1 & Acc. & F1 \\
\midrule
1024  
& 44.12 & 562 (4.23\%) & 0.9576 & 0.9665 & 0.9272 & 0.9834 & 0.9020 & 0.9924 & 0.9921 \\
4096  
& 54.45 & 698 (5.25\%) & 0.9473 & 0.9573 & 0.9093 & 0.9801 & 0.9038 & 0.9911 & 0.9907 \\
16384 
& 50.38 & 784 (5.90\%) & 0.9409 & 0.9514 & 0.9043 & 0.9773 & 0.8971 & 0.9874 & 0.9868 \\
\bottomrule
\end{tabular}
\end{table*}

\begin{table*}[t]
\centering
\caption{\small{MMoE Experimental Results across Byte Lengths (Average of Five Seeds).}}
\label{tab:mmoe_1d_avg}
\begin{tabular}{c c c c cc cc cc}
\toprule
\multirow{2}{*}{Byte Length} 
& \multirow{2}{*}{Test time} & \multirow{2}{*}{ASR} & \multirow{2}{*}{CDR}
& \multicolumn{2}{c}{Task 1}
& \multicolumn{2}{c}{Task 2}
& \multicolumn{2}{c}{Task 3} \\
\cmidrule(lr){5-6} \cmidrule(lr){7-8} \cmidrule(lr){9-10}
& (sec) & & & Acc. & F1 & Acc. & F1 & Acc. & F1 \\
\midrule
1024  
& 241.65 & 693 (5.22\%)  & 0.9478 & 0.9612 & 0.9205 & 0.9843 & 0.9641 & 0.9807 & 0.9075 \\
4096  
& 500.54 & 924 (6.96\%)  & 0.9304 & 0.9458 & 0.8988 & 0.9824 & 0.9627 & 0.9752 & 0.9051 \\
16384 
& 798.63 & 1018 (7.22\%) & 0.9277 & 0.9227 & 0.9130 & 0.9721 & 0.9477 & 0.9646 & 0.8732 \\
\bottomrule
\end{tabular}
\end{table*}
\begin{figure*}[htbp]

    \centering
    \begin{subfigure}[b]{1\linewidth}
        \centering
        \includegraphics[width=0.8\linewidth]{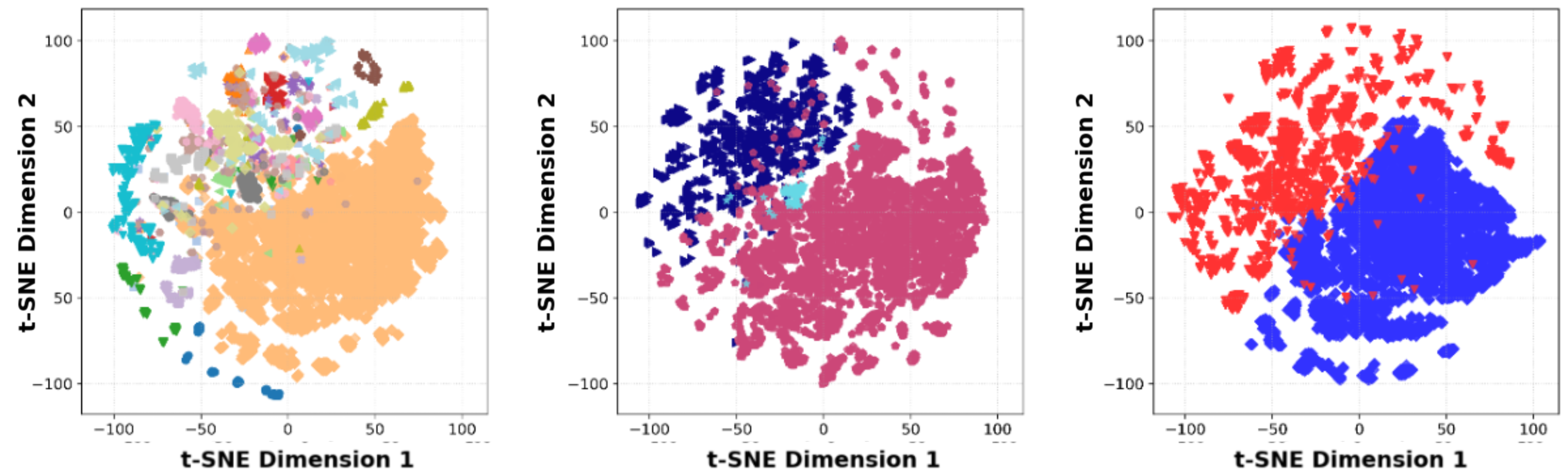}
        \caption{t-SNE visualization for the Multi-Gate Mixture-of-Experts (MMoE) framework trained on mutation-augmented Dataset D3 (Seed 750). These projections demonstrate the distinct functional specialization achieved across three tasks: Graph 1 depicts the malware family classification space, Graph 2 illustrates the packing status discrimination space, and Graph 3 represents the binary label detection space. Together, they reflect the highest Combined Detection Rate (CDR) and the most significant reduction in Attack Success Rate (ASR) by processing EMBER feature representations of Windows binaries.}
        \label{fig:mmoe_tsne}
    \end{subfigure}
    \vspace{0.1cm} 

    \begin{subfigure}[b]{1\linewidth}
        \centering
        \includegraphics[width=0.80\linewidth]{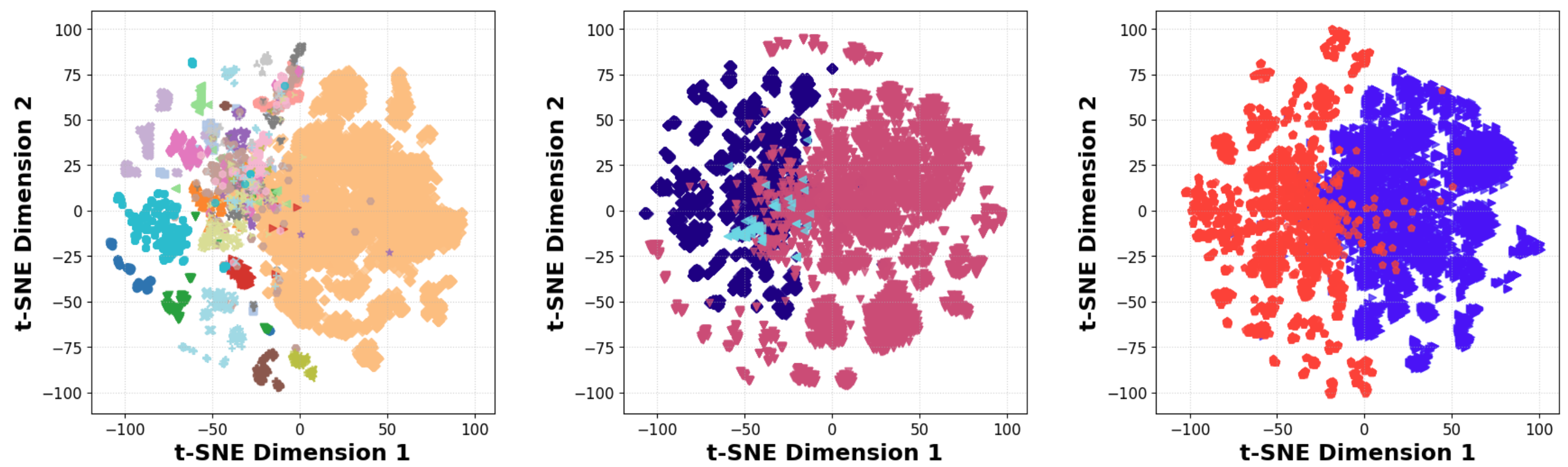}       
        \caption{Latent expert feature space t-SNE visualization for the Heterogeneous Mixture-of-Experts (HeteroMoE) framework, trained on Dataset D4 (Seed 750). This plot illustrates the optimal multi-task representation obtained from 1D image-based binary representations, which achieved better task discrimination than raw feature inputs.}
        \label{fig:hetero_tsne}
    \end{subfigure}
    \vspace{0.1cm} 

    \begin{subfigure}[b]{1\linewidth}
        \centering
        \includegraphics[width=0.80\linewidth]{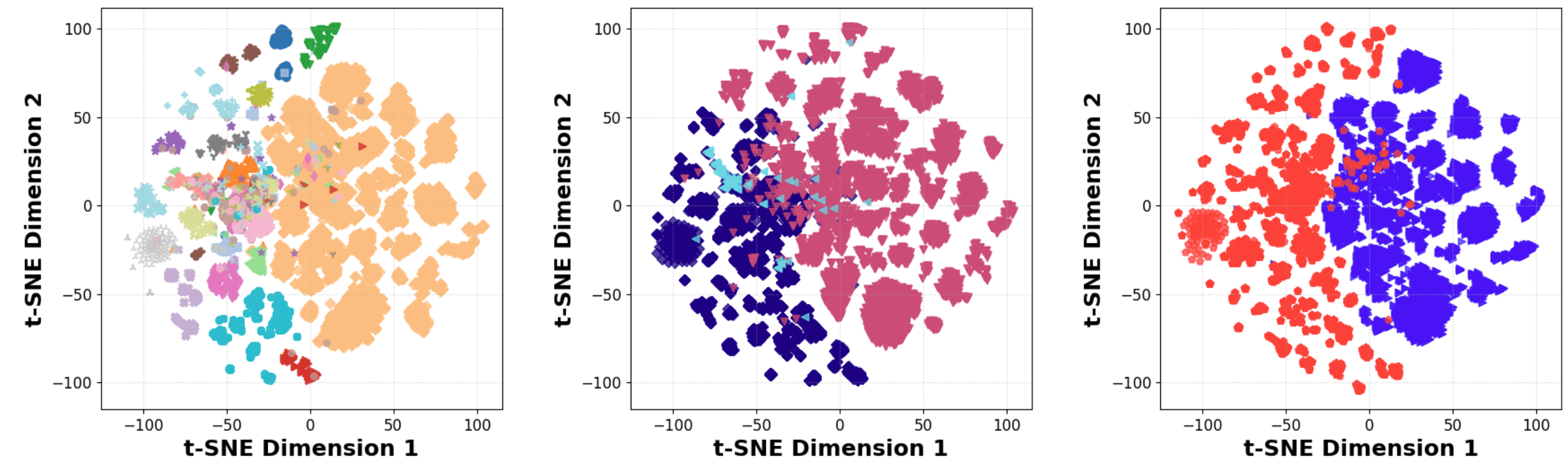}
        \caption{t-SNE visualization for the Homogeneous Mixture-of-Experts (HomoMoE) framework, trained on Dataset D4 (Seed 1024), showing the classifications across multiple tasks using 1D image representation. }
        \label{fig:homo_tsne}
    \end{subfigure}\par %
    \vspace{0.1cm} 
    \begin{subfigure}[b]{1\linewidth}
        \centering
        \includegraphics[width=0.85\linewidth]{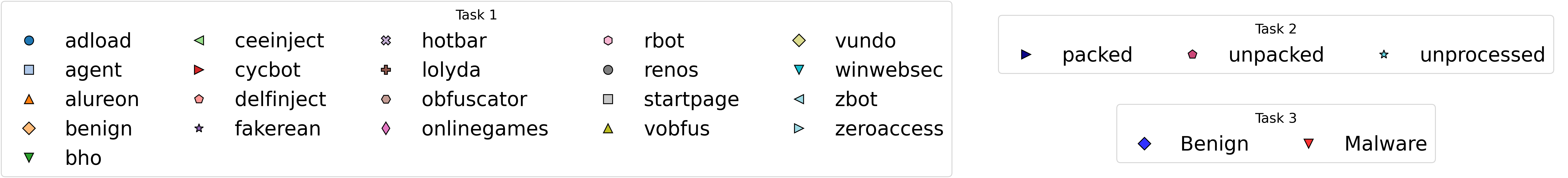}       
        \caption{Category mappings for Tasks 1, 2, and 3.}
        \label{fig:label chart}
    \end{subfigure}

    \caption{\small{The t-SNE projections on the experiments which gave the best results across MMoE, HeteroMoE, and HomoMoE configurations collectively illustrate the relationship between architectural design, input representation, and learned feature quality. MMoE, operating on EMBER features with Dataset D3 (Seed 750), achieves the highest CDR (0.9755) and lowest ASR (2.45\%), with its t-SNE projections reflecting well-separated task-specific gated spaces across family classification, packing status, and binary detection. HeteroMoE with input as 1D image-based binary representations at 1,024-byte length achieves a CDR of 0.9585 with moderate ASR ($4.14\%$), and it exhibits strong packing status separation and reasonable binary label isolation, driven by heterogeneous expert dimensionality. HomoMoE, also using 1D image inputs at $1,024$-byte length with CDR of $0.9576$ and ASR of 562 ($4.23\%$), demonstrates tighter malware family clustering relative to HeteroMoE, suggesting that uniform expert capacity better consolidates family-level structure from image-based representations. Together, these visualizations confirm that EMBER-based MMoE yields superior overall task discrimination, while image-based MoE variants trade global separability for more structured expert-level specialization.}}
    \label{fig:master_tsne_comparison}
\end{figure*}
\begin{figure*}[b]
    \centering
    \includegraphics[width=0.85\textwidth]{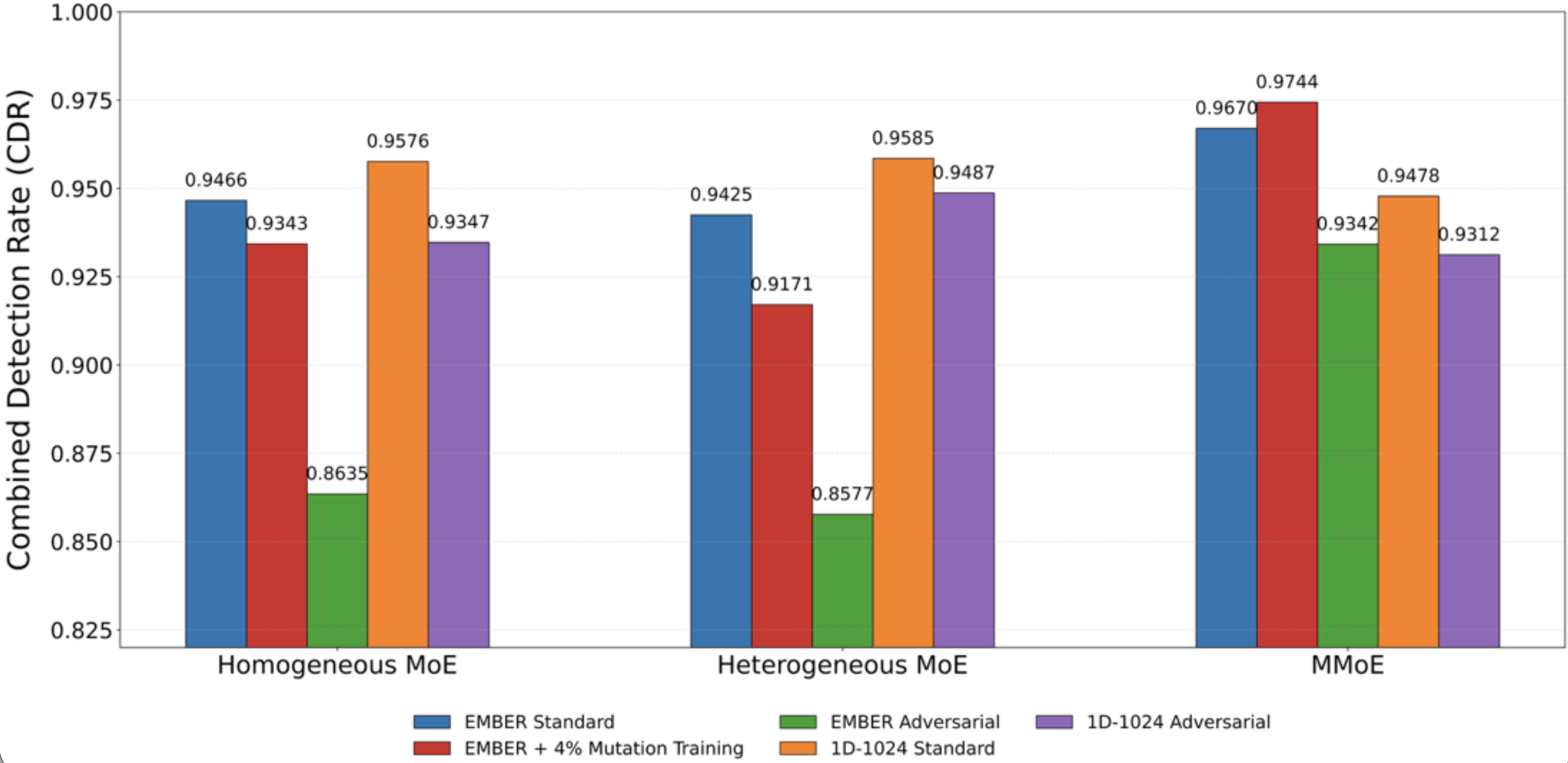}
    \caption{\small{Comparison of EMBER feature representations and 1D image representations (1024-byte length) across Homogeneous MoE, Heterogeneous MoE, and Multi-Gate MoE (MMoE) architectures under standard, mutation-augmented, and adversarial evaluation settings. Standard evaluation corresponds to training on 53,120 original samples and testing on 13,280 original test samples, while mutation-augmented evaluation incorporates mutated samples into the training set. Adversarial evaluation is performed using 1,392 mutated malware samples. Performance is measured using Combined Detection Rate (CDR), which requires simultaneous correct prediction across malware family classification, packed versus unpacked detection, and malware versus benign identification tasks. Among all configurations, MMoE using EMBER features achieves the highest CDR under both augmented training ($97.44\%$) and adversarial evaluation ($93.42\%$), while 1D image representations remain competitive, particularly for Homogeneous and Heterogeneous MoE frameworks under adversarial conditions.}}
    \label{fig:conclusion_comparison}
\end{figure*}

\begin{table*}[ht]
\centering
\caption{\small{Heterogeneous MoE -- 1D image (1024) Adversarial Attack Results.}}
\label{tab:hetero_1d_adv}
\begin{tabular}{c c c c cc cc c}
\toprule
\multirow{2}{*}{Seed} 
& \multirow{2}{*}{Test time} & \multirow{2}{*}{ASR} & \multirow{2}{*}{CDR}
& \multicolumn{2}{c}{Task 1}
& \multicolumn{2}{c}{Task 2}
& \multicolumn{1}{c}{Task 3} \\
\cmidrule(lr){5-6} \cmidrule(lr){7-8} \cmidrule(lr){9-9}
& (sec) & & & Acc. & F1 & Acc. & F1 & Acc. \\
\midrule
42   
& 15.35 & 63 (4.53\%) & 0.9547 & 0.9640 & 0.9183 & 0.9806 & 0.9515 & 0.9942 \\
123  
& 14.02 & 75 (5.39\%) & 0.9461 & 0.9568 & 0.9052 & 0.9849 & 0.9395 & 0.9985 \\
500  
& 13.71 & 77 (5.54\%) & 0.9446 & 0.9590 & 0.9098 & 0.9777 & 0.9403 & 0.9949 \\
750  
& 16.13 & 74 (5.32\%) & 0.9468 & 0.9568 & 0.9058 & 0.9798 & 0.9416 & 0.9964 \\
1000 
& 17.49 & 68 (4.89\%) & 0.9511 & 0.9626 & 0.9157 & 0.9770 & 0.9537 & 0.9949 \\
\midrule
Avg. ($\mu$) 
& 15.34 & 71 (5.11\%) & 0.9487 & 0.9598 & 0.9110 & 0.9800 & 0.9453 & 0.9958 \\
Std.Dev. ($\sigma$) 
& $\pm$1.44 & $\pm$5.16 (0.37\%) & $\pm$0.0037 & $\pm$0.0030 & $\pm$0.0052 & $\pm$0.0028 & $\pm$0.0060 & $\pm$0.0015 \\
\bottomrule
\end{tabular}
\end{table*}

\begin{table*}[ht]
\centering
\caption{\small{Homogeneous MoE -- 1D image (1024) Adversarial Attack Results.}}
\label{tab:homo_1d_adv}
\begin{tabular}{c c c c cc cc c}
\toprule
\multirow{2}{*}{Seed} 
& \multirow{2}{*}{Test time} & \multirow{2}{*}{ASR} & \multirow{2}{*}{CDR}
& \multicolumn{2}{c}{Task 1}
& \multicolumn{2}{c}{Task 2}
& \multicolumn{1}{c}{Task 3} \\
\cmidrule(lr){5-6} \cmidrule(lr){7-8} \cmidrule(lr){9-9}
& (sec) & & & Acc. & F1 & Acc. & F1 & Acc. \\
\midrule
42   
& 46.13 & 85 (6.11\%) & 0.9389 & 0.9439 & 0.9042 & 0.9849 & 0.9562 & 0.9945 \\
123  
& 49.84 & 93 (6.68\%) & 0.9331 & 0.9418 & 0.9038 & 0.9773 & 0.9335 & 0.9938 \\
500  
& 44.57 & 99 (7.11\%) & 0.9288 & 0.9389 & 0.8890 & 0.9820 & 0.9425 & 0.9941 \\
750  
& 51.22 & 87 (6.25\%) & 0.9375 & 0.9461 & 0.9064 & 0.9841 & 0.9345 & 0.9948 \\
1000 
& 47.90 & 90 (6.47\%) & 0.9353 & 0.9446 & 0.8980 & 0.9827 & 0.9417 & 0.9943 \\
\midrule
Avg. ($\mu$) 
& 47.93 & 91 (6.52\%) & 0.9347 & 0.9431 & 0.9003 & 0.9822 & 0.9417 & 0.9943 \\
Std.Dev. ($\sigma$) 
& $\pm$2.30 & $\pm$5.39 (0.39\%) & $\pm$0.0039 & $\pm$0.0027 & $\pm$0.0070 & $\pm$0.0030 & $\pm$0.0093 & $\pm$0.0004 \\
\bottomrule
\end{tabular}
\end{table*}

\begin{table*}[ht]
\centering
\caption{\small{Multi-Gate MoE (MMoE) -- 1D image (1024) Adversarial Attack Results.}}
\label{tab:mmoe_1d_adv}
\begin{tabular}{c c c c cc cc c}
\toprule
\multirow{2}{*}{Seed} 
& \multirow{2}{*}{Test time} & \multirow{2}{*}{ASR} & \multirow{2}{*}{CDR}
& \multicolumn{2}{c}{Task 1}
& \multicolumn{2}{c}{Task 2}
& \multicolumn{1}{c}{Task 3} \\
\cmidrule(lr){5-6} \cmidrule(lr){7-8} \cmidrule(lr){9-9}
& (sec) & & & Acc. & F1 & Acc. & F1 & Acc. \\
\midrule
42   
& 89.54 & 88 (6.32\%)  & 0.9368 & 0.9504 & 0.9043 & 0.9770 & 0.9371 & 0.9935 \\
123  
& 56.32 & 89 (6.40\%)  & 0.9361 & 0.9476 & 0.9005 & 0.9756 & 0.9382 & 0.9957 \\
500  
& 76.65 & 120 (8.62\%) & 0.9138 & 0.9418 & 0.8922 & 0.9619 & 0.9025 & 0.9835 \\
750  
& 87.54 & 85 (6.11\%)  & 0.9389 & 0.9533 & 0.9019 & 0.9777 & 0.9388 & 0.9943 \\
1000 
& 76.67 & 97 (6.97\%)  & 0.9303 & 0.9483 & 0.8988 & 0.9720 & 0.9287 & 0.9899 \\
\midrule
Avg. ($\mu$) 
& 77.344 & 96 (6.88\%) & 0.9312 & 0.9483 & 0.8995 & 0.9728 & 0.9291 & 0.9914 \\
Std.Dev. ($\sigma$) 
& $\pm$11.86 & $\pm$12.5 (0.89\%) & $\pm$0.0089 & $\pm$0.0039 & $\pm$0.0041 & $\pm$0.0058 & $\pm$0.0137 & $\pm$0.0044 \\
\bottomrule
\end{tabular}
\end{table*}

\subsubsection{Adversarial Evaluation of Raw 1D Image Representations on Dataset D5}
Testing the models against the mutation-based adversarial variants in dataset D5 exposes the limitations and expected performance degradation across all models. The Heterogeneous MoE framework (Table~\ref{tab:hetero_1d_adv}) maintains the strongest adversarial defense, achieving an average CDR of $0.9487$. This represents a minor drop from its standard baseline performance ($\text{CDR} = 0.9585$ on D4). Conversely, the Homogeneous MoE (Table~\ref{tab:homo_1d_adv}) and the Multi-Gate Mixture-of-Experts (MMoE) framework (Table~\ref{tab:mmoe_1d_adv}) exhibit greater sensitivity to the mutations, restricting their average CDRs to $0.9347$ and $0.9312$, respectively. As a result, the Attack Success Rate (ASR) is limited by the Heterogeneous MoE to $5.11\%$, which is less than the Homogeneous MoE ($\text{ASR} = 6.52\%$) and the MMoE framework ($\text{ASR} = 6.88\%$).

Analyzing performance across individual tasks, for Task 1, the Heterogeneous MoE achieves an average F1-score of $0.9110$, followed by the Homogeneous MoE and MMoE with scores of $0.9003$ and $0.8995$, respectively. The Heterogeneous MoE records the highest F1-score of $0.9453$ on Task 2, with the Homogeneous MoE and MMoE achieving $0.9417$ and $0.9291$, respectively. On Task 3, all three models demonstrate consistent performance, with average accuracy ranging from $0.9914$ to $0.9958$.

\begin{table}[ht]
\centering
\small
\framebox[\linewidth]{
\begin{minipage}{0.96\linewidth}
\vspace{2pt}
\textbf{Summary:} Under adversarial evaluation on 1D image representation(D5), the Heterogeneous MoE demonstrates the strongest robustness, achieving the highest CDR of $0.9487$ and leading F1-scores across Task 1 ($0.9110$) and Task 2 ($0.9453$), while all three models maintain consistent performance on Task 3.
\vspace{2pt}
\end{minipage}
}
\end{table}

\section{Discussion}
\label{sec:discussion}
This section explains the test results and how the models behave. Figure \ref{fig:conclusion_comparison} represents a comparison of the three models in different experiments.
Considering the best results out of the two types of feature representations used, the single-gate Homogeneous ($\text{CDR} = 0.9576$) and Heterogeneous ($\text{CDR} = 0.9585$) models show lower performance than the MMoE model ($\text{CDR} = 0.9744$, $\text{ASR} = 2.56\%$). This performance gap happens because of how they route information: single-gate models force one routing decision across different tasks, which creates conflict between the tasks. By giving each task its own gating network, MMoE lets them select experts independently without interfering with each other's learning. The ablation study shows that using a smaller $[256, 128]$ expert network with a task-tower size of $64$ gives the best result. Furthermore, the MMoE model demonstrates high computational efficiency, requiring a total test time of $0.0272\text{ seconds}$ for $13,280$ samples, which translates to a test time of $2.05\ \mu\text{s}$ per sample. The explainability analysis was conducted for the three models by plotting the t-SNE visualization for three tasks, which is illustrated in Figure \ref{fig:master_tsne_comparison}. The t-SNE graph of MMoE exhibits the most tightly packed isolated clusters, indicating that its gating network routes task-specific features into distinct, highly specialized expert representations with minimal overlap.

Changing the input from structured EMBER features to raw 1D image (Dataset D4) highlights the trade-offs between creating static features and letting the model learn directly from raw data. At a length of 1024 bytes, the Heterogeneous MoE works best, reaching a CDR of $0.9585$ ($\text{ASR} = 4.14\%$). On the other hand, the MMoE model drops in performance on raw sequence data ($\text{CDR} = 0.9478$). However, the results of MMoE using EMBER features as input are notably higher than the best results achieved using MoE models with a 1D image as input.

When tested against mutated adversarial samples, the single-gate models have a drop in performance when experimented with EMBER features as input, with Homogeneous and Heterogeneous CDRs decreasing to $0.8635$ and $0.8577$, respectively. This shows that changing a few metadata fields easily breaks their shared routing system and affects all tasks at the same time. When using a 1D image as input, the performance of MMoE is negligibly lower than that of the MoE frameworks. In contrast, MMoE performs comparatively better on EMBER features while performing adversarial attacks, attaining a CDR of $0.9342$ ($\text{ASR} = 6.58\%$).

\begin{figure}
\centering
\includegraphics[width=0.5\columnwidth]{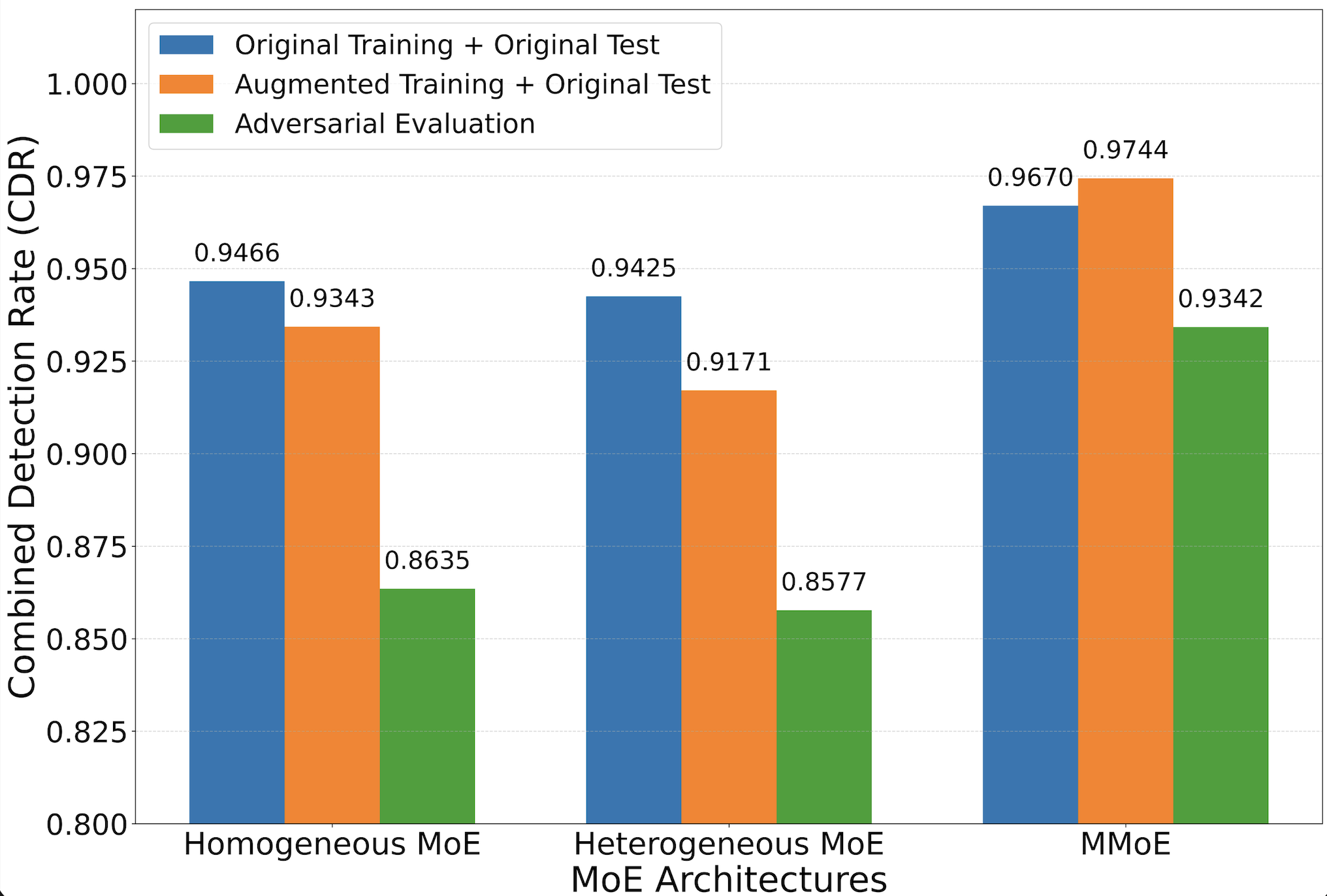}
\caption{\small{Comparison of Mixture of Experts (MoE) architectures under different evaluation settings using EMBER feature representations. The first setting corresponds to training on $53,120$ original samples and evaluating on $13,280$ original test samples. The second setting incorporates $4\%$ mutation-based augmented samples into the training set and evaluates the models on the same $13,280$ original test samples. The third setting evaluates adversarial robustness using 1,392 mutated malware samples. The graph presents the average Combined Detection Rate (CDR) achieved by the Homogeneous MoE, Heterogeneous MoE, and Multi-Gate MoE (MMoE) frameworks across the three experimental settings.}}
\label{fig:performance}
\end{figure}                                                                  

\begin{figure}
\centering
\includegraphics[width=0.5\columnwidth]{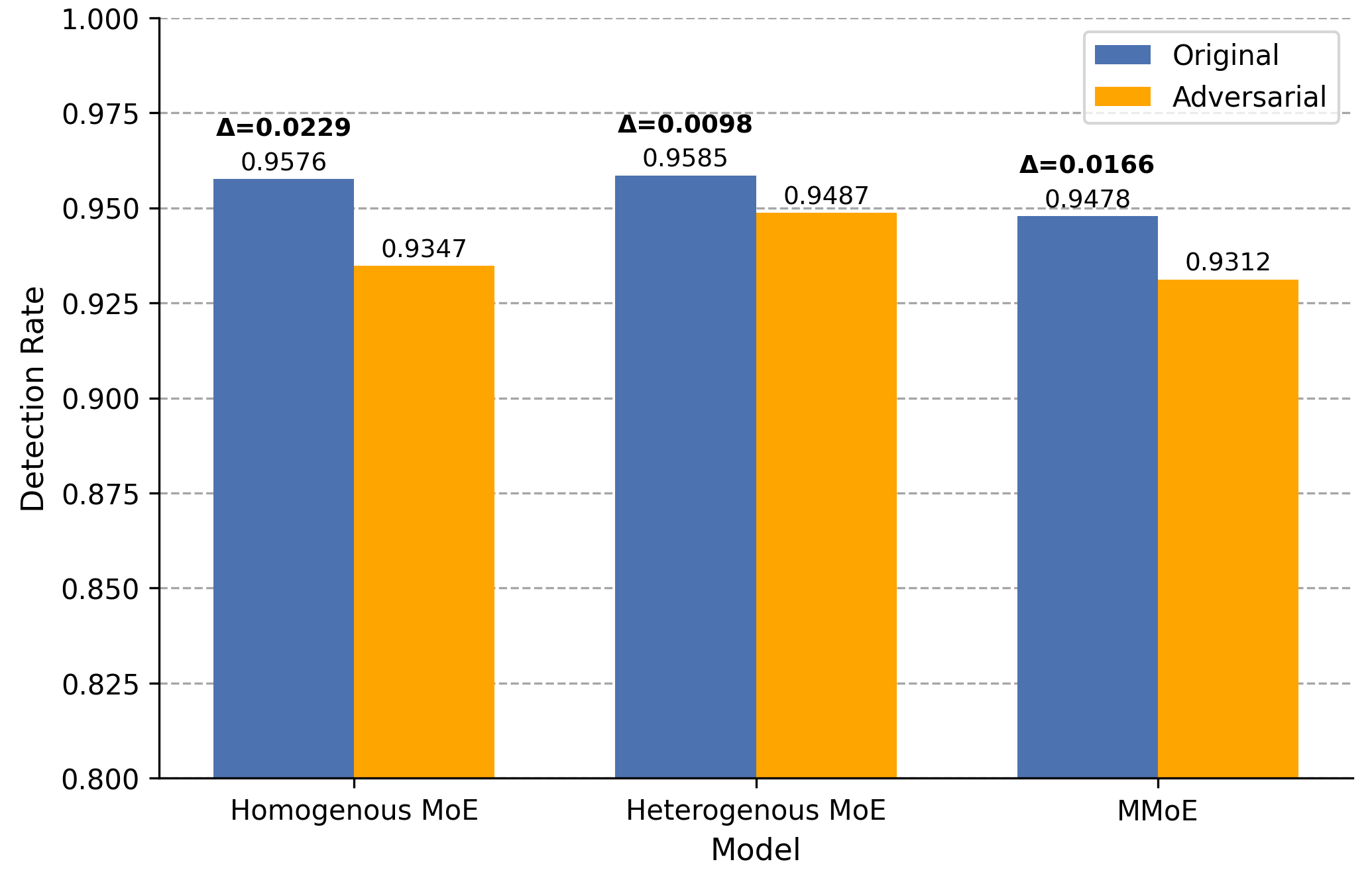}
\caption{\small{Performance Analysis of Mixture of Experts (MoE) models using 1D image representations at a length of 1024 bytes under standard and adversarial conditions. The first configuration uses raw sequential inputs to track the standard baseline performance on dataset D4. In the second approach, 1,392 structurally altered malware samples are used to test the models' adversarial robustness. The graph displays the average Combined Detection Rate (CDR) attained by the Homogeneous MoE, Heterogeneous MoE, and Multi-Gate MoE (MMoE) frameworks. (The absolute performance loss from the original to adversarial conditions is explicitly quantified by $\Delta$).}}
\label{fig:byte_performance_1024}
\end{figure}

\section{Conclusion and Future Work}
\label{sec:conclusion}
This research introduces a multi-task learning framework based on Mixture of Experts (MoE) architectures to handle malware classification, packing detection, and family attribution simultaneously. The system was evaluated using both high-dimensional EMBER features and raw 1D images under standard and mutation-based adversarial conditions. Our experimental evaluation shows how using task-specific routing weights helps the framework learn better representations across distinct security tasks, reducing the conflict often found in single-gate architectures. The framework maintains reliable classification stability even when processing mutated samples. These results suggest its suitability for real-world malware analysis.
Moreover, the obtained results show that MMoE exhibits the highest performance when using structured EMBER features, especially with mutation-augmented training and adversarial evaluation. By contrast, when it comes to raw 1D byte-level representations, the single-gate MoE variants and, in particular, the Heterogeneous MoE, remain more competitive. This demonstrates that the advantage obtained with task-specific gating is not independent of the input representation, but it is related to the stability and semantic richness of the extracted features.

The future plans include integrating Explainable AI (XAI) tools to explain model reasoning and analyze the specific features driving expert selection. We will investigate the impact of image size on model performance and experiment with different and larger datasets to test the generalizability of the framework. We also plan to transform the raw binary arrays into 2D image representations to evaluate how vision-based deep learning methods perform within the multi-task setup. In addition, we aim to experiment with the early fusion of EMBER and 1D features to analyze their combined effectiveness in improving the performance. Furthermore, we will explore the robustness of MoE architectures on adversarial examples crafted leveraging diverse metamorphic engines like SecML.

\section*{Data Availability}

The datasets used in this study are publicly available from the following sources:

Malware samples were obtained from MalwareBazaar (\url{https://bazaar.abuse.ch/}).  
Benign samples were collected from PortableApps (\url{https://portableapps.com/}) and the Practical Security Analytics dataset (\url{https://www.practicalsecurityanalytics.com}).  

The processed dataset and feature representations can be made available upon request.

\bibliographystyle{plain}
\bibliography{biblio}

\end{document}